\documentstyle[preprint,aps,epsf]{revtex}
\newcommand{\pspicture}[1]{
\centerline{\setlength\epsfxsize{9.2cm}\epsfbox{#1}}}
\ifx\DeclareFontShape\undefined
    \newcommand{\mathbf}[1]{{\bf #1}}
    \newcommand{\mathrm}[1]{{\rm #1}}
    \newcommand{\mathcal}[1]{{\cal #1}}
    \newcommand{\mathsf}[1]{{\sf #1}}
\else
    \DeclareSymbolFont{lasy}{U}{lasy}{m}{n}
    \SetSymbolFont{lasy}{bold}{U}{lasy}{b}{n}
    \let\Box\undefined
    \DeclareMathSymbol\Box{0}{lasy}{"32}
\fi

\ifx\epsffile\undefined
\else
\fi

\begin{document}
\title{%
A lattice calculation of the
branching ratio for some of the exclusive modes of $b \to s\gamma$.}
\author{{\it \large UKQCD Collaboration%
\footnote{kcb@th.ph.ed.ac.uk, jflynn@southampton.ac.uk
}}}
\address{ }
\author{K.C.~Bowler,  N.M.~Hazel, D.S.~Henty, H.~Hoeber, R.D.~Kenway,
D.G.~Richards, H.P.~Shanahan, J.N.~Simone}
\address{Department of Physics, The University of Edinburgh,
Edinburgh EH9~3JZ, Scotland}
\author{J.M.~Flynn, B.J.~Gough}
\address{Department of Physics, University of Southampton, Southampton
SO17~1BJ, UK}
\date{hep-lat/9407013, February 1995}
\maketitle
\begin{abstract}
We calculate the leading--order matrix element for
exclusive decays of $b \to s \gamma$
in the quenched approximation of
lattice QCD on a $24^3\times48$ lattice at $\beta{=}6.2$, using an
$O(a)$-improved fermion action. The matrix element is used to extract
the on-shell form factor $T_1(q^2{=}0)$ for $B \to K^*\gamma$ and
$B_s \to \phi \gamma$, using two different assumptions about the $q^2$
dependence of the form factors for these decays. For $B \to K^*\gamma$,
$T_1(q^2{=}0)$ is determined
to be $0.159^{+34}_{-33}\pm 0.067$ or $0.124^{+20}_{-18} \pm 0.022$
in the two cases.
We find the results to be consistent (in the
Standard Model) with the CLEO experimental branching ratio of
$\mbox{\it{BR\,}}(B\to K^*\gamma) = (4.5 \pm 1.5 \pm 0.9) \times 10^{-5}$.
\end{abstract}
\widetext

\section{Introduction}
\subsection{The Standard Model and New Physics}
Theoretical interest in the rare decay $B \to K^*\gamma$ as a test of
the Standard Model has been renewed by the experimental results of the
CLEO collaboration~\cite{cleo:evidence-for-penguins}. For the first
time, this mode has been positively identified and a preliminary
determination of its branching ratio given.

The radiative decays of the $B$ meson are remarkable for several
reasons.  The decay $B \to K^*\gamma$ arises from the
flavour-changing quark-level process $b \to s\gamma$, which occurs
through penguin diagrams at one-loop in the Standard Model.  As a
result, the decay is a purely quantum effect and a subtle
test of the Standard Model.  The process is also sensitive to new
physics appearing through virtual particles in the internal
loops. Existing bounds on the $b \to s\gamma$ branching ratio have
been used to place constraints on supersymmetry (SUSY)~%
\cite{bertolini:supersymmetry,%
oshimo:supersymmetry,%
barbieri:supersymmetry,%
lopez:supersymmetry,%
garisto:supersymmetry,%
diaz:supersymmetry,%
borzumati:supersymmetry}
and other extensions of the Standard Model (SM)~%
\cite{rizzo:two-higgs-doublet,%
hou:fourth-generation}.
A comprehensive review of these results can be found
in~\cite{hewett:top-ten}. Finally, it is also remarkable that this
rare process has a sufficiently large branching ratio to be detected
experimentally.  Thus, {\it accurate\/} experimental measurements and
{\it accurate\/} theoretical calculations of these decays could soon
probe new physics at comparatively low energies.

In order to compare the experimental branching ratio with a
theoretical prediction it is necessary to know the relevant hadronic
matrix elements. These have been estimated using a wide range of
methods, including relativistic and nonrelativistic quark models~%
\cite{deshpande:rel-quark-model,%
odonnel:rel-quark-model,%
altomari:nonrel-quark-model},
two-point and three-point QCD sum rules~%
\cite{dominguez:2pt-sum-rules,%
aliev:2pt-sum-rules,%
ball:3pt-sum-rules,%
colangelo:3pt-sum-rules-plb,%
ali:3pt-sum-rules,%
narison:3pt-sum-rules}
and heavy quark symmetry~\cite{ali:heavy-quark-symmetry}, but there
remains some disagreement between the different results. It is
therefore of interest to perform a direct calculation of the matrix
elements using lattice QCD\@. The viability of the lattice approach
was first demonstrated by the work of Bernard, Hsieh and
Soni~\cite{bhs:lattice-91} in 1991.

\begin{figure}
\pspicture{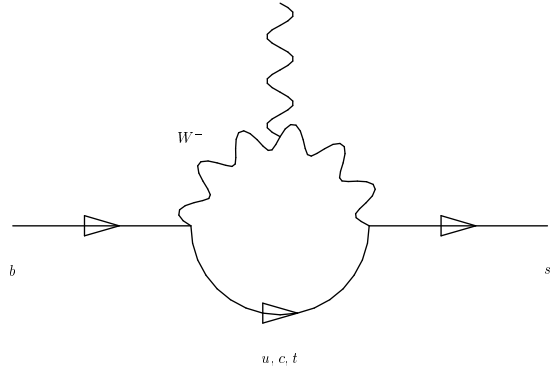}
\caption{An example of a penguin diagram contributing to the
decay $b\to s\gamma$.}
\label{figure:bsgwloop}
\end{figure}

Excluding QCD contributions, the free quark decay $b \to s\gamma$ in the
SM proceeds by diagrams similar to that shown in~Fig.(\ref{figure:bsgwloop}).
The
charm and top quark dominate, because the up quark contribution to the
loop is suppressed by the small CKM factor~$|V^{\phantom{*}}_{ub}
V^*_{us}|$.

If the value of the top mass is assumed, the Standard Model can be
tested by deriving an independent result for $\mbox{\it{BR\,}}(B \to
K^*\gamma)$. Deviations from the expected branching ratio would be an
indication of contributions to the decay from physics beyond the SM,
to which this decay is potentially sensitive.

Research on such contributions can be classified into supersymmetric
and non-supersymmetric extensions of the SM\@. In the latter case, Cho
and Misiak~%
\cite{cho_misiak_left_right}
considered $SU(2)_L \otimes SU(2)_R$ left-right symmetric models and
found considerable variations from the SM result for a wide range of
the free parameters, while Randall and Sundrum~%
\cite{randall_technicolour}
found significant potential deviations from the SM in technicolour
models.  Anomalous $WW\gamma$ couplings in $b \to s\gamma$ have been
analysed and the results found to be consistent with the SM\@. The
bounds obtained from this approach can improve on those from direct
searches~%
\cite{%
chia:anomalous-WWg,%
peterson:anomalous-WWg,%
rizzo:anomalous-WWg,%
he:anomalous-WWg}.
The contributions from two Higgs doublet models~%
\cite{glashow:typeII-two-higgs-doublet,%
abbot:typeI-two-higgs-doublet}
have been analysed to obtain bounds on the charged Higgs mass and
$\tan{\beta}$, the ratio of the vacuum expectation values of the
doublets~\cite{buras:review,hewett:two-higgs-doublet}.

SUSY models also involve additional Higgs doublets, but the
contribution of other boson-fermion loops, in particular charginos
($\chi^-$) with up type squarks, and gluinos ($\tilde{g}$) or
neutralinos ($\chi^0$) with down type squarks must also be included~%
\cite{bertolini:supersymmetry,%
oshimo:supersymmetry,%
barbieri:supersymmetry,%
lopez:supersymmetry,%
garisto:supersymmetry,%
diaz:supersymmetry,%
borzumati:supersymmetry,%
diaz:supersymmetry-ii}.
A thorough study of the decay in the Minimal Supersymmetric
Standard model can be found in
reference~\cite{borzumati:supersymmetry}. There are a strong
contributions from chargino and gluino loops, especially for large
$\tan{\beta}$, which interfere destructively with the Higgs
contribution and allow SUSY to mimic the SM in some regions of
parameter space.  As a result, the current limits on $\tan{\beta}$ and
Higgs masses are weak, but will tighten as more stringent bounds on
superpartner masses are obtained.

For the rest of this paper, we shall use the SM as the appropriate
model, and look for possible deviations from the experimental
branching ratio. It should be noted that the lattice calculation is
needed only to determine the effects of low energy QCD, and these are
independent of new physics. The low energy effect of many extensions
of the SM will be completely contained within the renormalisation
group operator coefficients, and hence it is straightforward to allow
for contributions from different models.

\subsection{Exclusive vs. Inclusive decay modes}

The inclusive decay $B \to X_s \gamma$ is predominantly a short distance
process and can be treated perturbatively in the spectator
approximation.  It is also possible to use Heavy Quark Effective
Theory (HQET) to compute the leading $1/m_b^2$
corrections~\cite{falk:B-to-Xs-gamma}.
The experimental inclusive branching ratio has been determined at
CLEO~\cite{cleo:inclusive},
\begin{equation}
\mbox{\it{BR\,}}(B \to X_s \gamma)=
(2.32 \pm 0.51 \pm 0.29 \pm 0.32)\times 10^{-4}.
\end{equation}
The procedure for obtaining this result has a mild model dependence
(the final result is a function of $m_b$).

In addition, the branching ratios of the exclusive decay modes of
$b \to s\gamma$ can also be experimentally determined, and the present
published branching ratio for $B \to K^*\gamma$ from the CLEO
collaboration~\cite{cleo:evidence-for-penguins} is,
\begin{equation}
\label{eq:CLEO_BR}
\mbox{\it{BR\,}}(B \to K^*\gamma) =  (4.5 \pm 1.5 \pm 0.9) \times 10^{-5}.
\end{equation}
The advantage of this measurement is that the experimental number is
model independent. Theoretical calculation of the relevant exclusive
matrix element requires the determination of long distance QCD
contributions which cannot be determined perturbatively, but can be
computed using lattice QCD.

\subsection{The Effective Hamiltonian and Hadronic Matrix Elements}
In order to determine the low energy QCD contributions to this decay,
the high energy degrees of freedom must be integrated out, generating
an effective $\Delta B = -1$, $\Delta S = 1$ hamiltonian.  Grinstein,
Springer and Wise~\cite{grinstein:b-meson-decay} determined the
hamiltonian ${\cal H}_{\mbox{\scriptsize\it eff}}$, to leading order in
weak matrix elements,
\begin{equation}
{\cal{H}}_{\mbox{\scriptsize\it eff}} =
-\frac{4 G_F}{\sqrt{2}} V_{tb}^{\phantom{*}} V_{ts}^*
\sum_{i=1}^8 C_i(\mu) O_i ,
\end{equation}
where the eight operators $O_i$ are multiplied by the renormalisation
group coefficients, $C_i(\mu)$. Six of the operators are four-quark
operators and two are magnetic moment operators, coupling to the gluon
and photon~\cite{simma:effective-lagrangians}.  The operator which
mediates the $b \to s\gamma$ transition is,
\begin{equation}
O_7=\frac{e}{16 \pi^2} m_b \overline{s} \sigma_{\mu\nu}
\frac{1}{2}(1+\gamma_5)b~F^{\mu\nu}.
\end{equation}
The coefficients $C_i(\mu)$ are set by matching to the full theory at
the scale $\mu = M_W$. The coefficient $C_7(m_b)$ is determined using
the renormalization group to run down to the appropriate physical
scale $\mu = m_b$~\cite{ciuchini:btoxs}, and is approximately given by,
\begin{equation}
\label{eq:csevenmb}
C_7(m_b)  =  \eta^{-16/23} \left( C_7(M_W)  +
\frac{58}{135}(\eta^{10/23}  -  1)  +  \frac{29}{189}(
\eta^{28/23}  -  1) \right),
\qquad
\eta=\frac{\alpha_s(m_b)}{\alpha_s(M_W)},
\end{equation}
where, in the Standard Model \cite{cho:weak-hamiltonian},
\begin{equation}
\label{eq:csevenmW}
C^{SM}_7(M_W)  =  \frac{1}{2} \frac{x}{(x -1)^3} \left(
\frac{2}{3} x^2  +  \frac{5}{12}x  -  \frac{7}{12}
 -  \frac{x}{2} \frac{(3x - 2)}{(x-1)} \log{x} \right),
\qquad
x  =  \frac{m_t^2}{M_W^2}.
\end{equation}
The effects of scale uncertainty in the leading order approximation
have been considered by Buras {\it et al.}~\cite{buras:review}.

To leading order, the on-shell matrix for $B \to K^*\gamma$ is given by,
\begin{equation}
{\cal M}= \frac{e G_F m_b}{2 \sqrt{2} \pi^2}
C_7(m_b) V^{\phantom{*}}_{tb} V_{ts}^*  \eta^{\mu*} \langle K^* |
J_\mu
 | B \rangle ,
\end{equation}
where,
\begin{equation}
J_\mu = \overline{s} \sigma_{\mu\nu} q^\nu b_R,
\end{equation}
and $\eta$ and $q$ are the polarization and momentum of the emitted
photon.  As outlined by Bernard, Hsieh and Soni~\cite{bhs:lattice-91},
the matrix element $\langle K^* | \overline{s} \sigma_{\mu \nu}
q^\nu b_R | B \rangle$ can be parametrised by three form
factors,
\begin{equation}
\langle K^* | J_\mu | B \rangle  =  \sum_{i=1}^3 C^i_\mu T_i(q^2)  ,
\end{equation}
where,
\begin{eqnarray}
C^{1}_\mu & = &
2 \varepsilon_{\mu\nu\lambda\rho} \epsilon^\nu p^\lambda k^\rho, \\
C^{2}_\mu & = &
\epsilon_\mu(m_B^2 - m_{K^*}^2) - \epsilon\cdot q (p+k)_\mu, \\
C^{3}_\mu & = &
\epsilon\cdot q
\left( q_\mu - \frac{q^2}{m_B^2-m_{K^*}^2} (p+k)_\mu \right).
\end{eqnarray}
As the photon emitted is on-shell, the form factors need to be
evaluated at $q^2{=}0$.  In this limit,
\begin{equation}
T_2(q^2{=}0)  = -i T_1(q^2{=}0) ,
\label{eq:T1_T2_equal}
\end{equation}
and the coefficient of $T_3(q^2{=}0)$ is zero.  Hence, the branching
ratio can be expressed in terms of a
single form factor, for example,
\begin{equation}
\label{eq:decay_rate}
\mbox{\it{BR\,}}(B \to K^* \gamma )
 = \frac{\alpha}{8 \pi^4} m_b^2 G_F^2
            m_B^3 \tau_B \left(1-\frac{m_{K^\ast}^2}{m_B^2}\right)^3
| V^{\phantom{*}}_{tb} V_{ts}^* |^2 |C_7(m_b)|^2 |T_1(q^2{=}0)|^2.
\end{equation}
This paper concerns the evaluation of $T_1(0)$.
We shall outline how matrix elements of the form $\langle V | J_\mu
| P \rangle$, where $| P \rangle$ is a heavy-light pseudoscalar meson
and $| V \rangle$ is a strange-light vector meson, can be calculated in
lattice QCD and explain the computational details involved.  We shall
evaluate the form factors $T_1(q^2{=}0)$ and
$T_2(q^2{=}0)$, make some statements about the
systematic error, and compare the calculated value of
$\mbox{\it{BR\,}}(B \to K^*\gamma)$ with the results from CLEO.
\subsection{Heavy Quark Symmetry}
\label{hqs}
We cannot directly simulate $b$-quarks on the lattice, as will be
explained below. Instead, we calculate with a selection of quark
masses near the charm mass. This means that any results for the form
factors must be extrapolated to the $b$-quark scale.  Heavy quark
symmetry~\cite{isgur:form-factors} tells us that,
\begin{equation}\label{eq:hqs-scaling}
\begin{array}{rcl}
T_1(q^2_{max}) &\sim& m_P^{1/2} \\
T_2(q^2_{max}) &\sim& m_P^{-1/2}
\end{array}
\end{equation}
as the heavy quark mass, and hence the pseudoscalar meson mass, $m_P$,
grows infinitely large. Combining this with the relation $T_2(q^2{=}0) =
-iT_1(q^2{=}0)$ constrains the $q^2$ dependence of the form factors.

Pole dominance ideas suggest that,
\begin{equation}
T_i(q^2) = {T_i(0)\over (1 - q^2/m_i^2)^{n_i}}
\end{equation}
for $i=1,2$, where $m_i$ is a mass that is equal to $m_P$ plus $1/m_P$
corrections and $n_i$ is a power. Since $1-q^2_{max}/m_i^2 \sim 1/m_P$
for large $m_P$, the combination of heavy quark symmetry and the form
factor relation at $q^2=0$ implies that $n_1 = n_2 + 1$. Thus we could
fit $T_2(q^2)$ to a constant and $T_1(q^2)$ to a single pole form or
fit $T_2(q^2)$ to a single pole and $T_1(q^2)$ to a double pole. These
two cases correspond to,
\begin{equation}
T_1(0) \sim \cases{m_P^{-1/2}&single pole\cr
                m_P^{-3/2}&double pole\cr}.
\end{equation}
As we will see, our data for $T_2(q^2)$ appear roughly constant in
$q^2$ when $m_P$ is around the charm scale, but have increasing
dependence on $q^2$ as the heavy quark mass increases. We will fit to
both constant and single pole behaviours for $T_2(q^2)$ below.

\section{Lattice Field Theory}
The hadronic matrix element $\langle V | J_\mu | P \rangle$ for the
$b \to s\gamma$ transition can be obtained from the correlator $\langle 0
| J^{V}_\rho(x) T_{\mu\nu}(y) J_P^{\dagger}(0) | 0 \rangle$, where
$J_P$ and $J^{V}_\rho$ are interpolating fields for the $P$ and $V$
mesons, consisting of a heavy quark, $h$, a light quark, $l$, and a
strange quark, $s$;
\begin{eqnarray}
J_{P}(x)       &=& \overline{l}(x) \gamma_5 h(x),     \\
J^{V}_\rho(x)  &=& \overline{l}(x) \gamma_\rho s(x),  \\
T_{\mu \nu}(y) &=& \overline{s}(y) \sigma_{\mu\nu} h(y).
\end{eqnarray}
The full matrix element
$ \langle V | \overline{s} \sigma_{\mu \nu} {1\over2}(1 + \gamma_5)
h | P \rangle$ can be derived using the Minkowski space
relation,
\begin{equation}
\gamma^5 \sigma^{\mu \nu}  =  \frac{i}{2} \varepsilon^{\mu \nu
\rho \lambda} \sigma_{\rho \lambda}.
\end{equation}
In Euclidean space, the correlator
$\langle 0 | J^{V}_\rho(x) T_{\mu\nu}(y) J_P^{\dagger}(0) | 0 \rangle$
can be computed numerically using the functional integral,
\begin{eqnarray}
\label{eq:funcintexpr}
\langle 0 | J^{V}_\rho(x) T_{\mu\nu}(y) J_P^\dagger(0) | 0 \rangle
& = & \frac{1}{Z}
\int {\cal D} A {\cal D} q {\cal D} {\bar q}~J_\rho^{V}(x) T_{\mu\nu}(y)
J_P^\dagger(0)
\exp(-S[A,q,{\bar q}]), \\
& = & \frac{1}{Z}
\int {\cal D} A~\mbox{Tr}
\left(\gamma_5 H(0,y)  \sigma_{\mu\nu} S(y,x) \gamma_\rho
L(x,0) \right)
\exp(-S_{\mbox{\scriptsize\it{eff}}}) ,
\label{eq:func-int-trace}
\end{eqnarray}
where  $S[A,q,{\bar q}]$ is the QCD action and  $S(y,x)$, $H(y,x)$,
$L(y,x)$ are the propagators from $x$ to $y$ for the $s$,
$h$ and $l$ quarks.
Working in momentum space, we calculate the three-point correlator,
\begin{eqnarray}
C^{3pt}_{\rho\mu\nu}(t,t_f,\vec{p},\vec{q}) & = &
\sum_{\vec{x},\vec{y}}
e^{i \vec{p}\cdot \vec{x}}
e^{- i \vec{q}\cdot \vec{y}}
\langle J_{P}(t_f,\vec{x}) T_{\mu\nu}(t,\vec{y})
 J^\dagger_{V\rho}(0)\rangle \\
& \mathop{\longrightarrow}\limits_{t,t_f - t,T\to\infty} &
\sum_{\epsilon}
\frac{Z_{P}}{2 E_{P}}
\frac{Z_{V}}{2 E_{V}}
e^{-E_{P} (t_f-t) } e^{-E_V t}
\epsilon_\rho
\langle  P(p)  | \overline{h} \sigma_{\mu\nu} s |
V(k,\epsilon) \rangle.
\end{eqnarray}
To obtain the matrix element $\langle P(p) | \overline{h}
\sigma_{\mu\nu} s | V(k) \rangle$, we take the ratio,
\begin{equation}
\label{eq:3pt-correlator}
C_{\rho\mu\nu}(t,t_f,\vec{p},\vec{q}) =
\frac{C^{3pt}_{\rho\mu\nu}(t,t_f,\vec{p},\vec{q})
}{C^{2pt}_{P}(t_f-t,\vec{p}) C^{2pt}_{V}(t,\vec{p}-\vec{q})
},
\end{equation}
where the two-point correlators are defined as,
\begin{eqnarray}
C^{2pt}_{P}(t,\vec{p})
& = &
\sum_{\vec{x}} e^{i\vec{p}\cdot\vec{x}}
  \langle J^\dagger_{P}(t,\vec{x}) J^{\vphantom{\dagger}}_{P}(0)
\rangle  \nonumber \\
& = &
\frac{Z^2_{P}}{2 E_{P}} \left( e^{-E_{P}t}+e^{-E_{P}(T-t)} \right) ,
\label{eq:pseudoscalar-two-point} \\
C^{2pt}_{V}(t,\vec{k})
&= &
-{\displaystyle
\left(\frac{1}{3}\right)}
 \sum_{\vec{x}}  e^{i\vec{k}\cdot\vec{x}}
  \langle J_{V\sigma}^\dagger(t,\vec{x})
  J^{\sigma}_{V}(0) \rangle \nonumber \\
& = &
\frac{Z^2_{V}}{2 E_{V}} \left( e^{-E_{V}t}+e^{-E_{V}(T-t)} \right).
\label{eq:vector-two-point}
\end{eqnarray}
By time reversal invariance and
assuming the three points in the correlators of
Eq.(\ref{eq:3pt-correlator}) are sufficiently separated in time,
a term proportional to the required matrix element dominates:
\begin{equation}
\label{eq:3pt-asymptotic}
C_{\rho\mu\nu}  \mathop{\longrightarrow}\limits_{t,t_f - t,T\to\infty}
 \frac{1}{Z_{P}Z_{V}} \sum_\epsilon
\epsilon_{\rho}
\langle V(k,\epsilon) | \overline{s} \sigma_{\mu\nu} h
 | P(p) \rangle + \dots,
\end{equation}
and $C_{\rho\mu\nu}$ approaches a plateau. The three-point correlator
is calculated in its time reversed form to allow the use of
previously calculated light propagators. The factors $Z_{P}$, $Z_V$
and the energies of the pseudoscalar and vector particles are obtained
{}from fits to the two-point Euclidean correlators.

In order to simulate this decay on a sufficiently finely spaced
lattice, vacuum polarisation effects were discarded and the gauge
field configurations generated in the quenched approximation.  The
decays $D \rightarrow K e \nu$, $D \rightarrow K^* e \nu$, and $D_s
\rightarrow \phi e \nu$ have been calculated in the quenched
approximation~\cite{lubicz:d-decay-ii,bernard:d-decay,UKQCD:d-decay}
and have been found to be in relatively good agreement with
experiment.  It is therefore quite plausible to assume that the
systematic error from the quenched approximation for this calculation
would be of a similar size.

The matrix element $\langle K^* | \overline{s} \sigma_{\mu \nu}
q^\nu b_R | B \rangle$ cannot be directly calculated as
realistic light quarks cannot be simulated owing to critical slowing
down in determining the propagator for small masses.
Instead light quarks are simulated at a number of masses approximately
that of the strange quark mass, and any result is extrapolated to the
chiral limit.  Furthermore, $b$ quarks cannot be simulated directly as
the $b$-quark mass is greater than the inverse lattice spacing
($2.73(5) \, \mbox{GeV}$), and the variation of the propagator would
occur over lengths smaller than the lattice spacing.  As a result,
heavy quarks are simulated with masses around the charm quark mass,
and the results extrapolated to $m_b$.  Hence, $\langle V | J_\mu | P
\rangle $ has to be calculated at a number of different light, strange
and heavy quark masses.

\section{Computational Details}

Sixty $SU(3)$ gauge configurations were generated in the quenched
approximation for a $24^3 \times 48$ lattice at $\beta=6.2$.  These
configurations were generated with periodic boundary conditions using
the hybrid over-relaxed algorithm, and the standard discretised gluon
action, defined in~\cite{luscher}.  The configurations were separated
by $400$ compound sweeps, starting at sweep number $2800$.  The
inverse lattice spacing was determined to be $2.73(5) \, \mbox{GeV}$,
by evaluating the string tension~\cite{ukqcd:string-tension}.  In
physical units, this corresponds to a spacing of approximately $0.07
\, \mbox{fm}$ and a spatial size of $1.68 \, \mbox{fm}$.  In order to
simulate heavy quarks whose masses are approaching the inverse lattice
spacing, the $O(a)$-improved fermion action of Sheikholeslami and
Wohlert~\cite{sw-action} (also referred to as the clover action) was
used.  This is defined as,
\begin{equation}
S^C_F  =  S^W_F - i\frac{\kappa}{2}\sum_{x,\mu,\nu}\bar{q}(x)
F_{\mu\nu}(x)\sigma_{\mu\nu}q(x)  ,
\end{equation}
where $S^W_F$ is the standard Wilson fermion
action~\cite{ukqcd:string-tension,wilson-paper} and $F_{\mu\nu}$ is a
lattice definition of the field strength tensor, which we take to be
the sum of the four untraced plaquettes in the $\mu\nu$ plane open at
the point $x$,
\begin{equation}
F_{\mu\nu}(x) =
 \frac{1}{4} \sum_{\mathord{%
\hbox{\vrule\vbox{\hrule width0.3em\kern0.3em\hrule}\vrule}}=1}^{4}
 \frac{1}{2i}
 \biggl[U_{\mathord{%
\hbox{\vrule\vbox{\hrule width0.3em\kern0.3em\hrule}\vrule}}\mu\nu}(x)
 - U_{\mathord{%
\hbox{\vrule\vbox{\hrule
width0.3em\kern0.3em\hrule}\vrule}}\mu\nu}^\dagger(x)\biggr].
\end{equation}
In using this action, all observables with fermion fields
$q,\overline{q}$ must be ``rotated'',
\begin{eqnarray}
q(x) \;\; &\rightarrow& \;\; \left(1 - \frac{1}{2}\stackrel{\rightarrow}
{\! \not \!\! \Delta} \right) q(x), \\ \nonumber
\overline{q}(x) \;\; &\rightarrow& \;\; \overline{q}(x)
\left(1 + \frac{1}{2}\stackrel{\leftarrow}
{\! \not \!\! \Delta} \right)   ,
\end{eqnarray}
where $\Delta_\mu$ is the discretised covariant derivative, operating
on the quark fields as,
\begin{eqnarray}
\stackrel{\rightarrow}{\Delta_\mu} \, q(x) \;\;& = & \;\;
\frac{1}{2}\left( U_\mu(x) q(x + \mu) \,-\, U^\dagger_\mu(x - \mu) q(x -
\mu) \right) , \\ \nonumber
\overline{q}(x) \, \stackrel{\leftarrow}{\Delta_\mu}  \;\;& = & \;\;
\frac{1}{2}\left( \overline{q} (x + \mu) U^\dagger_\mu(x)
\,-\, \overline{q}(x - \mu) U_\mu(x - \mu)
\right).
\end{eqnarray}
This action eliminates the tree level $O(ma)$-error of the Wilson
action~\cite{heatlie:clover-action}, which can be significant for
heavy quark systems~\cite{ukqcd:fP,ukqcd:charmonium}.

For each configuration, quark propagators were calculated using the
over--relaxed minimal residual algorithm with red--black
preconditioning for $\kappa = 0.14144$, $0.14226$ and $0.14262$, using
periodic boundary conditions in the spatial directions and
anti--periodic boundary conditions in the temporal direction.
Smearing was not used in the calculation of these light propagators.
The first two $\kappa$ values can be used to interpolate to the
strange quark mass which corresponds to $\kappa = 0.1419(1)$
\cite{ukqcd:strange-prd}.
The third $\kappa$ value, corresponding to a somewhat lighter quark,
was used in conjunction with the others in order to test the behaviour
of the data in the chiral limit.

Heavy propagators, for $\kappa_h = 0.121$, $0.125$, $0.129$ and
$0.133$, were evaluated using timeslice $24$ of some of the above
propagators as the source.
For $\kappa_h = 0.121$ and $0.129$, the propagators for all
of the light $\kappa$ values were used.
For $\kappa_h = 0.125$ and $0.133$, the propagators
for $\kappa = 0.14144$ and $0.14226$ were used.
To reduce excited state contamination,
these sources were smeared using the gauge invariant Jacobi
algorithm~\cite{ukqcd:smearing}, with an r.m.s.\ smearing radius of~5.2.
Because of memory limitations, these propagators were evaluated
only for timeslices 7~to~16 and 32~to~41.

Using these propagators, the three point correlators
were evaluated.
The spatial momentum ${\bf p}$ was chosen to be
$(0,0,0)$ or $(\pi/12,0,0)$
(the lowest unit of momentum in lattice units that can be injected).
All possible choices of ${\bf q}$ were
calculated such that the magnitude of the spatial momentum of the
vector meson ${\bf k}$ was less than $\sqrt{2} \pi/12$.  This is
because the signal of light hadrons degrades rapidly as
the momentum  is increased \cite{alamos:light-spectrum}.

In order to obtain $\langle V | \overline{s} \sigma_{\mu \nu} h | P
\rangle$, the decay constant and energy were
determined for the pseudoscalar of each heavy--light $\kappa$
combination and the vector of each possible light $\kappa$
combination, for all possible momenta used.  The process of extracting
these is well understood and has been discussed in detail
elsewhere~\cite{ukqcd:strange-prd}.  As the two point functions
are periodic, a correlator at a time $0 \le t \le 24$ was averaged
with the same correlator at $48 - t$ to improve the statistical
sample.  This ``folded'' data was fitted
to~Eq.(\ref{eq:pseudoscalar-two-point}) or~Eq.(\ref{eq:vector-two-point}) for
timeslices 15 to 23.  For both the two point and three point functions
we utilised the discrete symmetries $C$, $P$ and $T$ (folding)
wherever possible, in addition to averaging over equivalent momenta.
The statistical errors for all correlators were determined by the
bootstrap procedure~\cite{efron:bootstrap}, using 1000 bootstrap
subsamples from the original configurations. The finite
renormalization needed for the lattice--continuum matching of the
$\sigma_{\mu\nu}$ operator has been calculated
\cite{borrelli:improved-operators} but has a negligible effect here
($O(2\%)$) and was not included.  It introduces a small correction to
the branching ratio which is considered in the conclusions.

As outlined in the previous section, the weak matrix elements
$C_{\rho \mu\nu}$ were
extracted from the three point data and the fits to the two point
data.
Having divided out
the contributions from the two point amplitudes and energies,
the matrix element $\langle V | \overline{s} \sigma_{\mu \nu} h | P
\rangle$ was isolated.
These matrix elements were combined to determine the
form factors $T_1(q^2)$, $T_2(q^2_{max})$ and
$T_2(q^2)$.
Each form factor was extracted by a correlated fit to a constant
for timeslices 11, 12 and 13.

\section{Results}
The data for unphysical masses, and off-shell photons must be combined
to isolate the form factors and extrapolate to the physical regime.
It is clear from~Eq.(\ref{eq:T1_T2_equal})
and~Eq.(\ref{eq:decay_rate}) that the branching ratio can be evaluated
{}from $T_1(q^2{=}0;m_B;m_{K^*})$ or $T_2(q^2{=}0;m_B;m_{K^*})$.  As
demonstrated in a previous paper~\cite{ukqcd:penguin-prl}, the
evaluation of $T_1(q^2{=}0;m_P;m_{K^*})$ is relatively
straightforward, and $T_2(q^2{=}0;m_P;m_{K^*})$ can be determined in a
similar way. To test heavy quark scaling, we also extracted the form
factor $T_2$ at maximum recoil, where $q^2=q^2_{max}=(m_P-m_V)^2$, in
the same way as Bernard {\it et al.}~\cite{bhs:penguin-prl}. These
form factors were extrapolated to the physical mass $m_P=m_B$, and an
estimate of systematic errors in the extrapolation made by comparing
different methods.
\subsection{Extraction of form factors}
\subsubsection{$T_1(q^2)$}
\begin{figure}
\pspicture{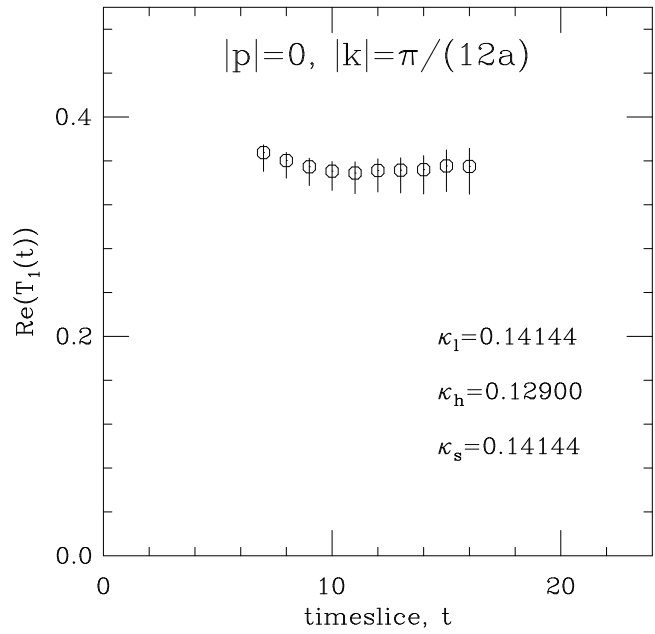}
\caption{A typical plot of $T_1(q^2{=}0;m_P;m_V)$ vs.\ time.
{}From the application of the time reversal operator, it can be
shown that only the real component of $T_1$ is non--zero.}
\label{figure:t1-vs-time}
\end{figure}

The form factor $T_1$ can be conveniently extracted from the matrix
elements by considering different components of the relation,
\begin{equation}
4( k^\alpha p^\beta -  p^\alpha k^\beta) T_1(q^2) =
\varepsilon^{\alpha\beta\rho\mu}
C_{\rho\mu\nu}q^{\nu}.
\end{equation}
We see a plateau in $T_1$ about $t=12$. The use of smeared operators
for the heavy quarks provides a very clean signal, with stable
plateaus forming before timeslice~11. The data for the heaviest of our
light quarks, $\kappa_l=\kappa_s=0.14144$, with the smallest
statistical errors, are shown in~Fig.(\ref{figure:t1-vs-time}).

\begin{figure}
\pspicture{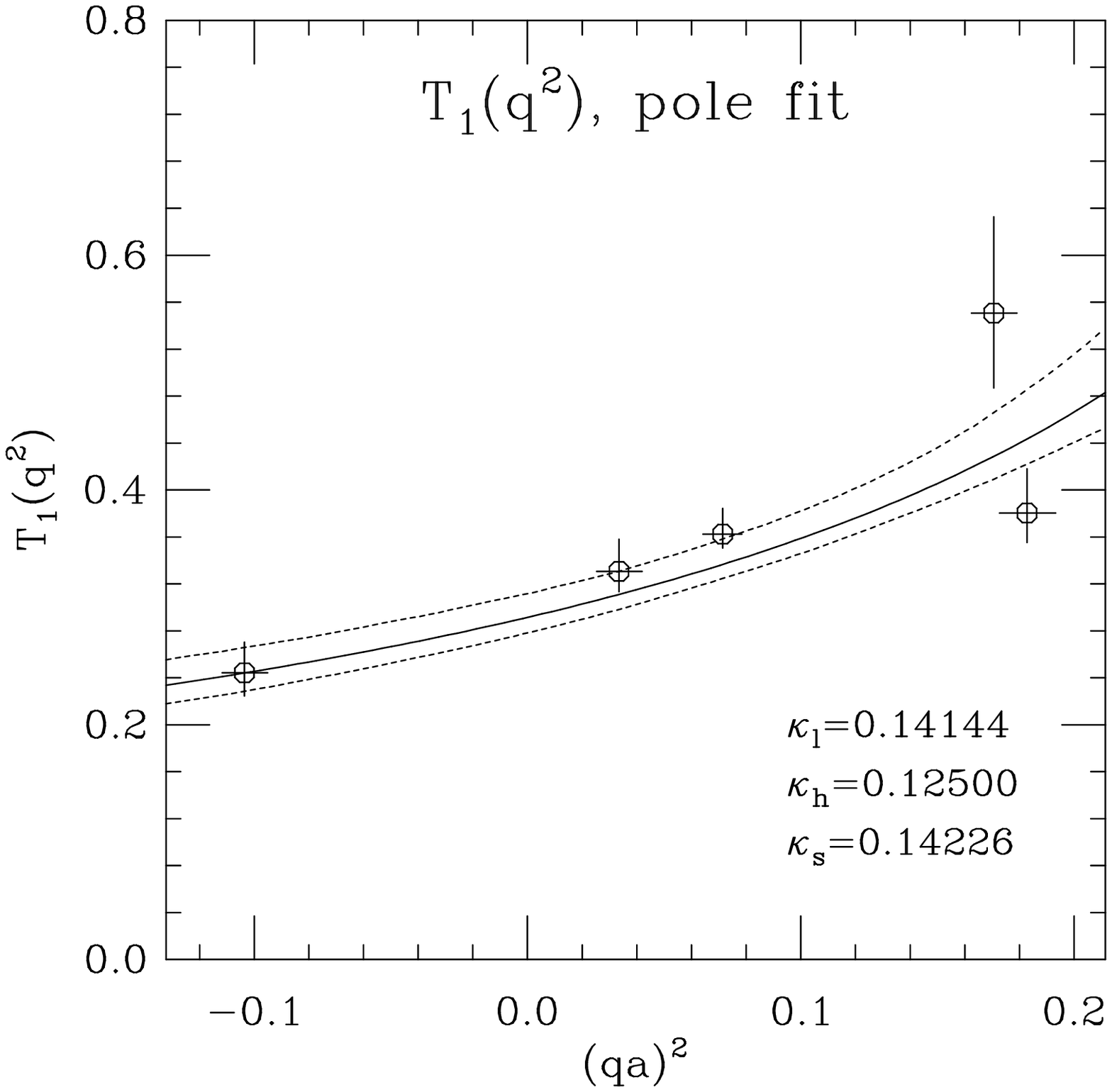}
\caption{$T_1(q^2)$, using a pole fit. The dotted lines represent
the 68\% confidence levels of the fit at each $q^2$.}
\label{figure:Tone_vs_qsq}
\end{figure}

The form factor is evaluated for each of the five possible values of
$q^2$.  We fit $T_1(q^2)$ to a pole or dipole model in order to obtain
the on-shell form factor $T_1(q^2{=}0)$,
\begin{equation}
T_1(q^2)= {T_1(q^2{=}0) \over 1- q^2/m^2},\qquad
T_1(q^2)= {T_1(q^2{=}0) \over (1- q^2/m^2)^2}.
\end{equation}
We allow for correlations between the energies of the vector and
pseudoscalar particles and $T_1$ at each $q^2$. An example of such a
fit, for $\kappa_l=\kappa_s=0.14144$, is shown
in~Fig.(\ref{figure:Tone_vs_qsq})
and the full set of fit parameters and their $\chi^2/\mbox{d.o.f.}$
are shown in tables~\ref{table:qsq_fits} and ~\ref{table:qsq_fits_b}.

\begin{figure}
\pspicture{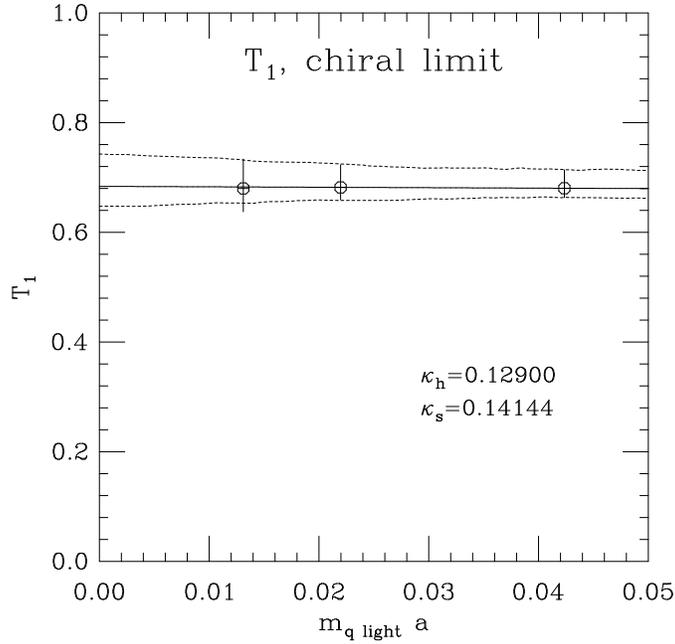}
\caption{Chiral extrapolation of $T_1(q^2{=}0)$.
The dotted lines indicate the 68\% confidence levels of the fit.
$a m_{q~light}$ is the lattice pole mass. }
\label{figure:t1-chiral-extrapolation}
\end{figure}

The chiral limit behaviour of $T_1(q^2{=}0;m_P;m_V)$, interpolated
{}from a single pole fit, was explored for $\kappa_h= 0.121$ and
$0.129$, in our earlier work~\cite{ukqcd:penguin-prl}. To
test for approximate spectator quark independence, we compared the
single pole fits of the form factor to the two functions,
\begin{eqnarray}
T_1(q^2{=}0;m_{q,\, light}) &=& a + b m_{l} ,\\
T_1(q^2{=}0;m_{q,\, light}) &=& c ,
\end{eqnarray}
where $m_{l}$ is the lattice pole mass,
\begin{equation}
m_l={1 \over 2}({1\over\kappa} - {1\over\kappa_{crit}}),
\end{equation}
and $\kappa_{crit}=0.14315(2)$~\cite{ukqcd:strange}.  The linear
coefficient $b$ was found to be consistent with zero for each
combination of $\kappa_s$ and $\kappa_h$ (see
Fig.\ref{figure:t1-chiral-extrapolation}).  From
table~\ref{table:Tone_chiral}, the $\chi^2/\mbox{d.o.f.}$ for both
fits are similar, indicating that for the data available, the
assumption that the form factor is a constant, independent of the
spectator quark mass, is valid.  Hence, the data for
$\kappa_l=0.14144$ was used for the chiral limit, and a simple linear
interpolation carried out between $\kappa_s=0.14144$ and $0.14226$ for
the strange quark, in order to obtain $T_1(q^2{=}0;m_P;m_{K^*})$.
These results are listed in the columns labelled (b) and (c) in
table~\ref{table:T1_T2_comparison}.
\subsubsection{$T_2(q^2)$}
\begin{figure}
\pspicture{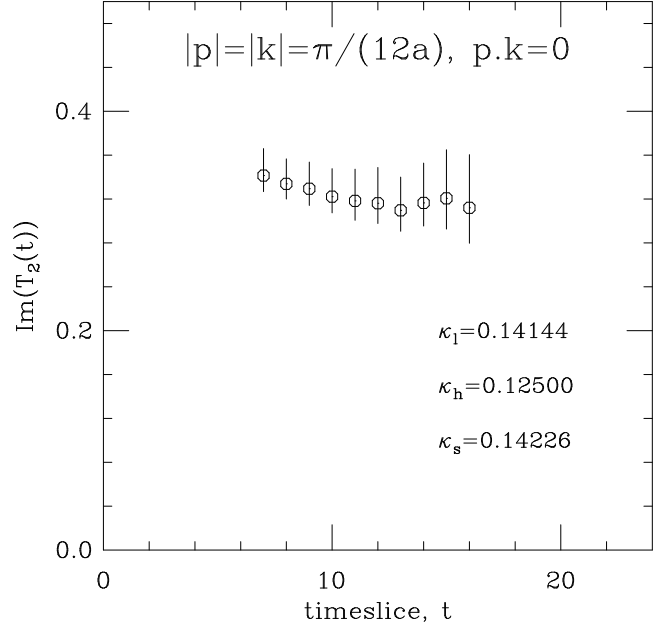}
\caption{$\mbox{Im}(T_2)$, for a typical momentum used.
{}From the application of the time reversal operator, it can be shown
that only the imaginary component of $T_2$ is non--zero.}
\label{figure:t2-vs-time}
\end{figure}

\begin{figure}
\pspicture{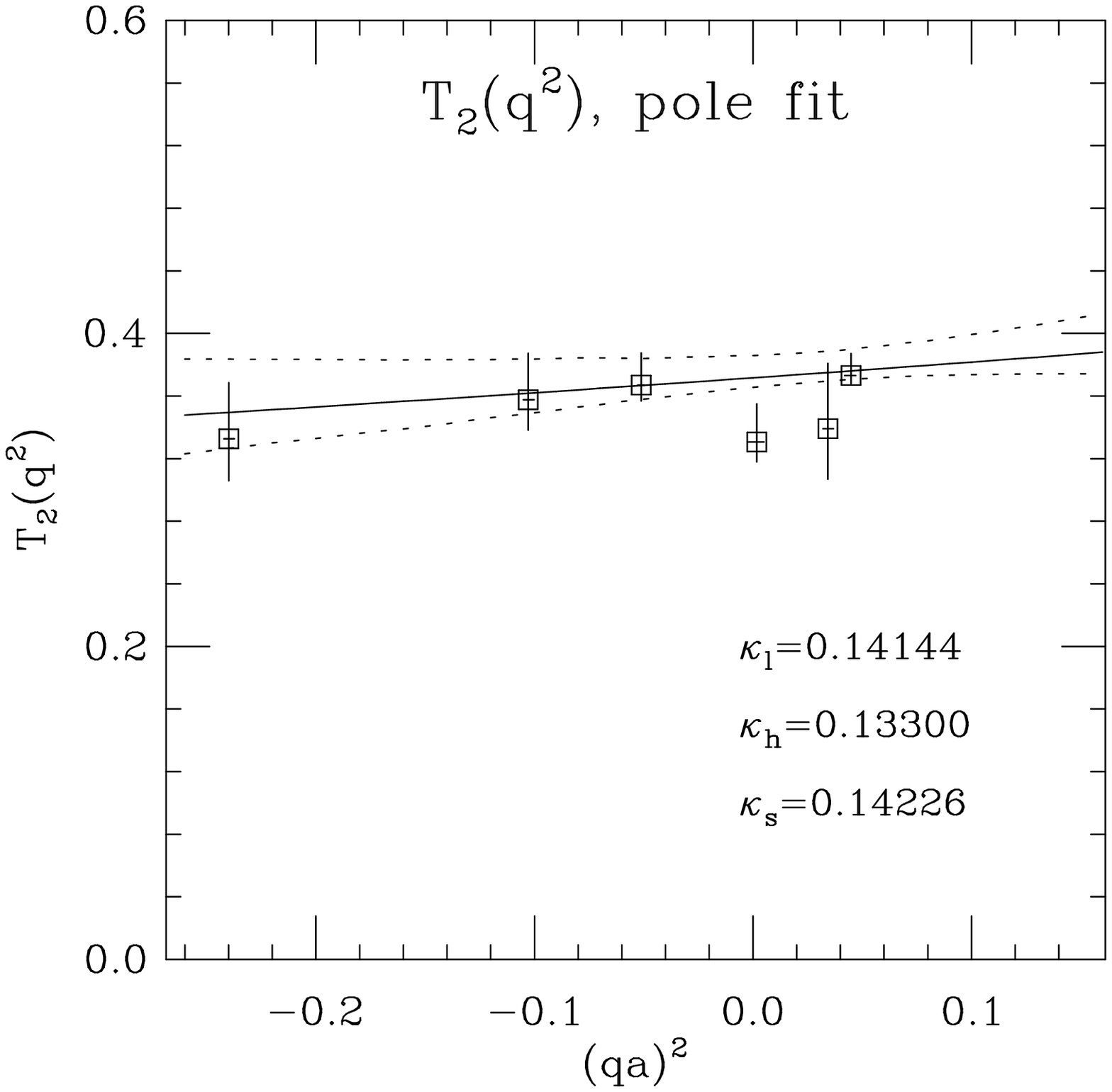}
\caption{$T_2(q^2)$, with a pole fit. The dotted lines represent
the 68\% confidence levels of the fit at each $q^2$.}
\label{figure:t2-vs-qsq}
\end{figure}

The form factor $T_2$ can be extracted from the matrix elements using
the same procedure as $T_1$, by considering the different components
of,
\begin{equation}
(m_P^2 - m_V^2) T_2(q^2;m_P;m_V) = C_{ii\nu} q^\nu,
\end{equation}
for all $i$ (not summed) such that $q^i=0$. A typical plateau for
$T_2$ is shown in~Fig.(\ref{figure:t2-vs-time}). We extract $T_2$ for a
range of $q^2$ as shown in~Fig.(\ref{figure:t2-vs-qsq}).

Fig.(\ref{figure:t2-vs-qsq}) shows that $T_2(q^2)$ is roughly constant
as a function of $q^2$ for our data, with heavy quark masses around
the charm mass.  We fit $T_2$ to a constant: we can then compare with
the value of $T_1(q^2=0)$ where $T_1$ is fitted with a single pole
form. We also fit $T_2$ to a single pole form (as shown in the figure)
and compare with $T_1(q^2=0)$ when $T_1$ is fitted with a double pole
form.  The results of the fits for $T_2$ are shown in
tables~\ref{table:T2_qsq_fits} and~\ref{table:T2_qsq_fits_b}, and the
chiral extrapolations for the single pole fit in
table~\ref{table:Two_inter_chiral}. The pole mass is found to be
large, and a linear behaviour holds well for all possible $q^2$,
including $q^2_{max}$, as shown in~Fig.(\ref{figure:t2-vs-qsq}). Once
again the data for $k_l = 0.14144$ was used for the chiral limit and
the results are listed in the columns labelled (d) and (e) in
table~\ref{table:T1_T2_comparison}.

\begin{figure}
\pspicture{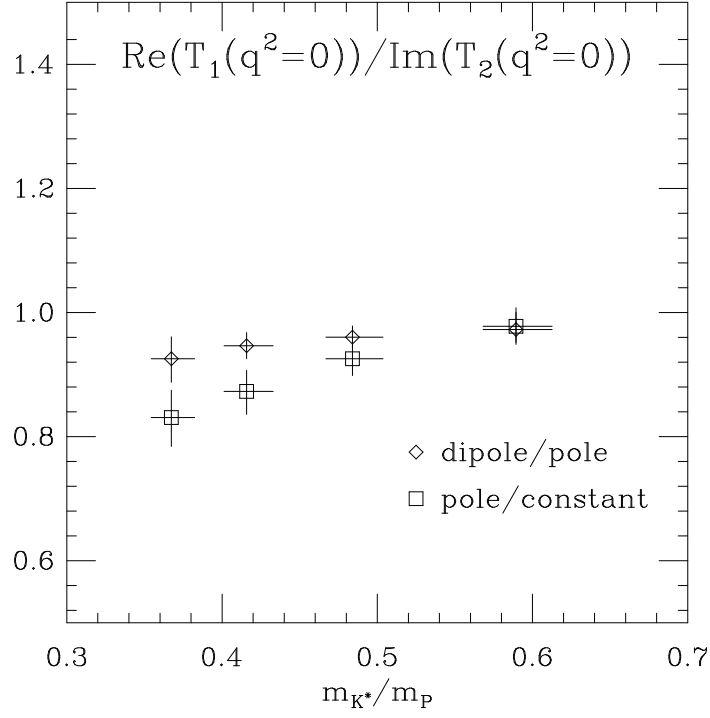}
\caption{The ratio $T_1/T_2$ at $q^2{=}0$ for dipole/pole and
pole/constant fits.}
\label{figure:T1_over_T2_comparison}
\end{figure}

The ratio $T_1(q^2{=}0;m_P;m_{K^*})/T_2(q^2{=}0;m_P;m_{K^*})$ is shown
in~Fig.(\ref{figure:T1_over_T2_comparison}). The two sets of points
show $T_1$ fitted to a double pole form and $T_2$ to a single pole or
$T_1$ fitted to a single pole and $T_2$ constant. The ratio should be
1, in accordance with the identity
$T_2(0)=-iT_1(0)$,~Eq.(\ref{eq:T1_T2_equal}).  We find greater
consistency from the double-pole/single-pole fit.

\subsubsection{$T_2(q^2_{max})$}
\begin{figure}
\pspicture{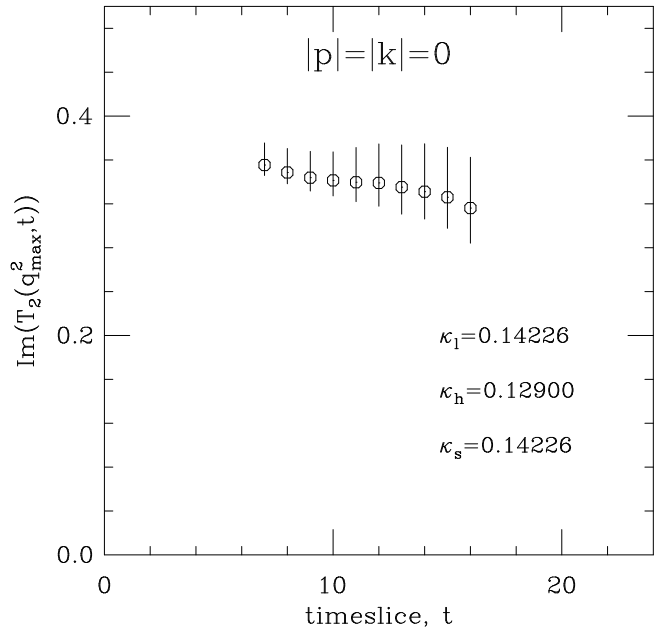}
\caption{A typical plot of $T_2(q^2_{max};m_P;m_V)$ vs.\ time.}
\label{figure:t2-qsq-max-vs-time}
\end{figure}

\begin{figure}
\pspicture{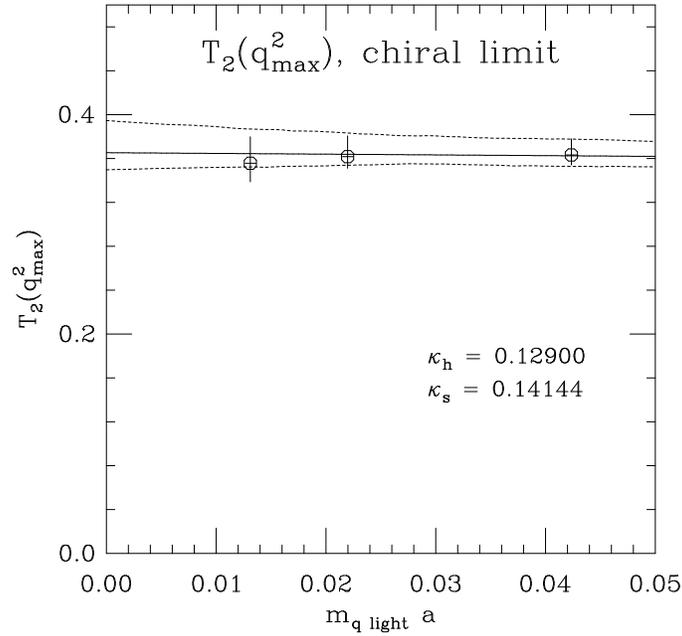}
\caption{Chiral extrapolation of $T_2(q^2_{max})$.}
\label{figure:T2_qsq_max_chiral}
\end{figure}

The evaluation of $T_2(q^2_{max};m_P;m_V)$ is also straightforward,
since at zero momentum, ${\bf p}{=}{\bf 0}$, ${\bf
k}{=}{\bf 0}$, the contributions from other form factors vanish,
\begin{eqnarray}
(m_P + m_V) \, T_2(q^2_{max})
	& = & C_{110}({\bf p}={\bf 0}, {\bf k}={\bf 0}),
\nonumber\\
	& = & C_{220}({\bf p}={\bf 0}, {\bf k}={\bf 0}),
 \nonumber\\
	& = & C_{330}({\bf p}={\bf 0}, {\bf k}={\bf 0}).
\end{eqnarray}
An example of this data is shown
in~Fig.(\ref{figure:t2-qsq-max-vs-time}).  The behaviour of
$T_2(q^2_{max};m_P;m_V)$ as a function of the spectator quark mass was
examined at $\kappa_h = 0.121$ and $0.129$ in the same way as for
$T_1(q^2{=}0)$.  It was again found that the linear coefficient $b$
was consistent with zero for each combination of $\kappa_s$ and
$\kappa_h$: see~Fig.(\ref{figure:T2_qsq_max_chiral}) for an
example. From table~\ref{table:Two_chiral}, the $\chi^2/\mbox{d.o.f.}$
for both fits are seen to be similar, indicating that for the data
available, the assumption that the form factor is independent of the
spectator quark mass is valid. Hence, the data for $\kappa_l=0.14144$
was used for the chiral limit, to obtain $T_2(q^2_{max};m_P;m_{K^*})$.

Bernard {\it et al.}~\cite{bhs:penguin-prl} converted this result to
$q^2{=}0$ by assuming single pole dominance,
\begin{equation}
\label{eq:pole_dominance}
T_2^{pole}(q^2) = \frac{T_2(0)}{1 - q^2/m_{P_{s1}}^2}.
\end{equation}
The current $J_\mu$ in the matrix element can be expressed in a $V +
A$ form, with $T_1$ corresponding to the vector component and $T_2$
and $T_3$ to the axial current.  Therefore, in a single pole model,
the exchanged particle, $P_{s1}$, for the $T_2$ form factor should be
the lowest $J^P = 1^+$ state with the correct spin, parity and
strangeness quantum numbers. We extracted
$T^{\mbox{\scriptsize\it{pole}}}_2(q^2{=}0;m_P;m_{K^*})$ from
$T_2(q^2_{max})$ using a single pole model, with the mass of the $1^+$
states determined from fits to two-point functions for each heavy
quark mass.  The results of these extrapolations are shown in the
column labelled (a) in table~\ref{table:T1_T2_comparison}.

\begin{figure}
\pspicture{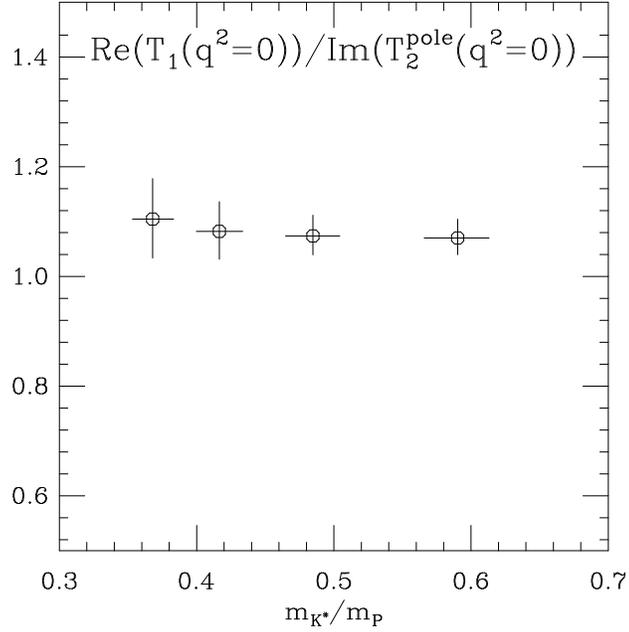}
\caption{The ratio
$T_1(q^2{=}0;m_P;m_{K^*})/T_2^{pole}(q^2{=}0;m_P;m_{K^*})$ with $T_1$
fitted to a dipole form and $T_2^{pole}$ extrapolated from
$T_2(q^2_{max}$) using a fitted $1^+$ mass.}
\label{figure:T1_on_T2_pole}
\end{figure}

The ratio $T_1(q^2{=}0;m_P;m_{K^*})/ T_2^{pole}(q^2{=}0;m_P;m_{K^*})$
is shown in~Fig.(\ref{figure:T1_on_T2_pole}).  We note that using a
fixed pole mass from two-point functions gives a 10-20\% difference in
the ratio (at the heaviest masses) compared with allowing the pole
mass to vary in the fits.


\subsection{Extrapolation to $M_B$}
The appropriate ansatz for extrapolating the on-shell form factor in
the heavy quark mass to $T_1(q^2{=}0;m_B;m_{K^*})$ is not {\it a
priori\/} clear. As we saw in section~\ref{hqs}, one has to model the
$q^2$ dependence of the form factors, maintaining consistency with
known heavy quark scaling results~\cite{isgur:form-factors} at
$q^2_{max}$, from~Eq.(\ref{eq:hqs-scaling}), and the relation
$T_1(0)=iT_2(0)$. Expanding unknown parameters in powers of $1/m_P$,
one obtains scaling laws for the on-shell form factors $T_1(q^2{=}0)$
and $T_2(q^2{=}0)$. Thus, while the scaling behaviour of
$T_2(q^2_{max})$ can be checked directly, the behaviours of $T_1(0)$
and $T_2(0)$ will depend on assumptions made for the $q^2$
dependence. We now address these issues.

Bernard {\it et al.}~\cite{bhs:penguin-prl} used the heavy-quark
scaling law for the off-shell form factor,
$T_2(q^2_{max};m_P;m_{K^*})$ to extrapolate $T_2$ to
$T_2(q^2_{max};m_B;m_{K^*})$, before applying a single pole dominance
model as before to reach the on-shell point
$T_2(q^2{=}0;m_B;m_{K^*})$. They estimated the appropriate pole mass.
The validity of the pole model over the wide range of momentum
transfer from $q^2{=}0$ to $q^2_{max}$ was required, but tests at
heavy quark masses around the charm quark mass showed it to be quite
accurate.

Our results for $T_2(q^2;m_P;m_{K^*})$,
see~Fig.(\ref{figure:t2-vs-qsq}), appear nearly independent of $q^2$
for masses $m_P$ around the charm scale. Hence, we have fitted $T_2$
to both single pole and constant forms, with corresponding behaviour
for $T_1$. This will give us two alternative forms for the heavy mass
dependence of $T_1(q^2{=}0;m_B;m_{K^*})$.

\subsubsection{$T_2(q^2_{max})$}
At $q^2{=}q^2_{max}$, the initial and final hadronic states have zero
spatial momentum and the contributions of form factors other than
$T_2$ vanish,
\begin{equation}
\label{eq:qsqmaxmatrixelement}
\langle K^* | \overline{s} \sigma_{\mu \nu} q^\nu b_R | B \rangle =
 \epsilon_\mu ( m_B^2 - m_{K^*}^2 ) T_2(q^2_{max}).
\end{equation}
In the heavy quark limit, the matrix element
of~Eq.(\ref{eq:qsqmaxmatrixelement}) scales as $m_B^{3/2}$, owing to the
normalisation of the heavy quark state ($\sqrt{m_B}$) and the momentum
transfer $q$ ($q^0=m_B-m_{K^*}$).  The leading term in the heavy quark
scaling of $T_2(q^2_{max})$ is expected to be $m_B^{-1/2}$, analogous
to the scaling of $f_B$\cite{neubert:HQET,ukqcd:fP}.  Higher order
$1/m_B$ and $1/m_B^2$ corrections will also be present, as will
radiative corrections~\cite{shifman:radiative-corrections,%
wise:radiative-corrections}.

\begin{figure}
\pspicture{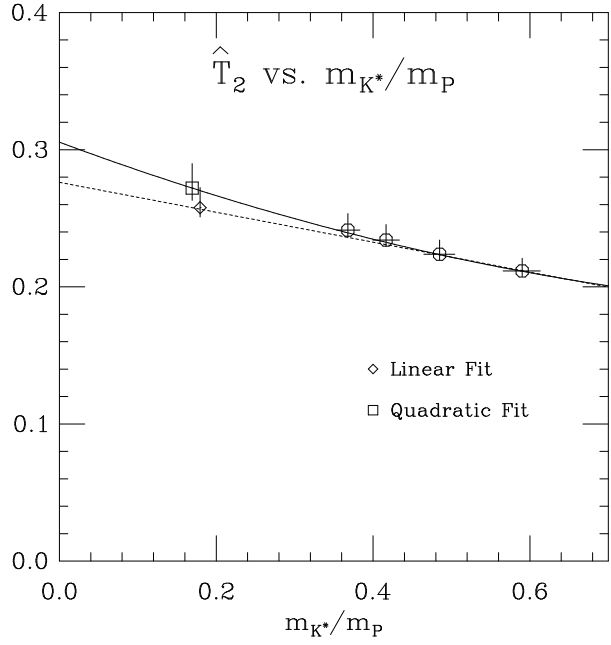}
\caption{Extrapolation of $T_2(q^2_{max})$ to $m_B$, assuming
HQET. The quantity plotted is $\hat T_2$, defined in the text, which
is equal to $T_2(q^2_{max})$ at $m_B$.}
\label{figure:T2_scaling}
\end{figure}

Hence, the form factor $T_2(q^2_{max})$ should scale as,
\begin{equation}
\label{eq:T2_scaling_law}
T_2(q^2_{max};m_P;m_{K^*}) \sqrt{m_P}
=
\mbox{const.} \times [ \alpha_s(m_P) ]^{-2/\beta_0}
 \left(1 + \frac{a_1}{m_P} + \frac{a_2}{m_P^2} + \dots\right).
\end{equation}
To test heavy quark scaling, we form the quantity,
\begin{equation}
{\hat T}_2=
T_2(q^2_{max})
\sqrt{m_P \over m_B}
\left({\alpha_s(m_P)\over\alpha_s(m_B)}\right)^{2/\beta_0},
\end{equation}
where we approximate $\alpha_s(\mu)$ by,
\begin{equation}
\alpha_s(\mu) = {2 \pi \over \beta_0 \ln( \mu/\Lambda_{QCD} ) }.
\end{equation}
with $\Lambda_{QCD}=200$ MeV and $\beta_0=11-{2\over3}N_f$. In the
quenched approximation, $N_f$ is taken to be zero.
The normalisation ensures that ${\hat T}_2=T_2(q^2_{max})$ at the
physical mass $m_B$. Linear and quadratic correlated fits
to~Eq.(\ref{eq:T2_scaling_law}) were carried out with the functions,
\begin{eqnarray}
{\hat T}_2(m_P)&=&A\left(1+{B\over m_P}\right), \\
{\hat T}_2(m_P)&=&A\left(1+{B\over m_P}+{C\over m_P^2}\right),
\end{eqnarray}
and are shown in~Fig.(\ref{figure:T2_scaling}).
Taking the quadratic fit of $T_2$ at $m_P = m_B$ as the best
estimate, and the difference between the central values of the
linear and quadratic fits as an estimate of the sytematic error, $T_2$
was found to be
\begin{equation}
\label{eq:Ttwo-qsqmax-result}
T_2(q^2_{max};m_B;m_{K^*}) = 0.269^{+17}_{-9}\pm{0.011} .
\end{equation}
%

Once $T_2(q^2_{max})$ is extracted, we can obtain $T_2(0)$ in the two
cases, pole model or constant, for the $q^2$ behaviour. If $T_2$ is
constant, then~Eq.(\ref{eq:Ttwo-qsqmax-result}) is the result at $q^2=0$.
In the pole model, the expected exchange particle for $T_2$ is the
$1^+$ $B_{s1}$ state, but experimental data for its mass is not yet
available. However, it is possible to estimate reasonable upper and
lower bounds for the mass from HQET\@.  It can be shown
that~\cite{neubert:HQET},
\begin{eqnarray}
\label{eq:mbs1mbsplitting}
m_{B_{s1}} - m_{B} &=& \Delta \overline{\Lambda} + \frac{A}{m_b} +
O(\frac{1}{m_b^2}), \\
m_{D_{s1}} - m_{D} &=& \Delta \overline{\Lambda} + \frac{A}{m_c} +
O(\frac{1}{m_c^2}).
\end{eqnarray}
Neglecting terms of order $1/m_c^2$, the upper and lower bounds
for~Eq.(\ref{eq:mbs1mbsplitting}) are,
\begin{equation}
\frac{m_c}{m_b}(m_{Ds1} - m_{D})
< m_{B_{s1}} - m_{B} <
m_{D_{s1}} - m_{D}
\end{equation}
Making the approximation,
\begin{equation}
\frac{m_c}{m_b} \sim  \frac{m_D + 3 m_{D^*}}{m_B + 3 m_{B^*}}
\end{equation}
the range of the expected pole mass can be found,
\begin{equation}
m_{B_{s1}} = 5.74\pm{0.21}~\mbox{GeV}.
\end{equation}
Therefore,
\begin{equation}
\label{eq:T2_calc}
T^{\mbox{\scriptsize\it{pole}}}_2(q^2{=}0;m_B;m_{K^*}) =
0.112^{+7}_{-7}\mbox{}^{+16}_{-15},
\end{equation}
where the first error is statistical and the second is the systematic
error obtained by combining the variation of the pole mass within its
bounds and the systematic error from~Eq.(\ref{eq:Ttwo-qsqmax-result}).
There is clearly a significant systematic difference between the
results in~Eq.(\ref{eq:Ttwo-qsqmax-result}) and~Eq.(\ref{eq:T2_calc})
corresponding to the two assumed forms for $T_2(q^2)$.

\subsubsection{$T_1(q^2{=}0)$}
If constant-in-$q^2$ behaviour is assumed for $T_2$, then $T_2(0)$
should satisfy the same scaling law as $T_2(q^2_{max})$
in~Eq.(\ref{eq:T2_scaling_law}). Combining this with the identity
$T_1(0)=iT_2(0)$ leads to a scaling law for $T_1(0)$:
\begin{equation}
\label{eq:T1_one_half_scaling}
T_1(0;m_P;m_{K^*}) \sqrt{m_P}
=
\mbox{const.} \times [ \alpha_s(m_P) ]^{-2/\beta_0}
 \left(1 + \frac{a_1}{m_P} + \frac{a_2}{m_P^2} + \dots\right).
\end{equation}

If single pole dominance is assumed for $T_2$ and the mass of
the exchanged $1^+$ particle can be expanded as,
\begin{equation}
\label{eq:mass-expansion}
m_{P_{s1}}=m_P \left(1 + {b_1 \over m_P} + {b_2 \over m_P^2} + \dots
\right),
\end{equation}
then $T_1(q^2{=}0;m_P;m_{K^*})$ should satisfy a modified scaling law,
\begin{equation}
\label{eq:T1_three_halves_scaling}
T_1(0;m_P;m_{K^*}) \, m_P^{3/2}
=
\mbox{const.} \times [ \alpha_s(m_P) ]^{-2/\beta_0}
 \left(1 + \frac{c_1}{m_P} + \frac{c_2}{m_P^2} + \dots\right),
\end{equation}
where the unknown coefficients in~Eq.(\ref{eq:mass-expansion}) have
been absorbed into the unknown scaling coefficients of the matrix
element.
A similar scaling relationship has been found by
Ali {\it et al.} \cite{ali:3pt-sum-rules} by the sum rules approach.

The two scaling forms were tested in the same way as for
$T_2(q^2_{max})$, by forming the quantities,
\begin{equation}
{\hat T}_1=
T_1(q^2{=}0)
\left({m_P \over m_B}\right)^{N/2}
\left({\alpha_s(m_P)\over\alpha_s(m_B)}\right)^{2/\beta_0}.
\end{equation}
where $N$ is 1 or 3 as appropriate.

\begin{figure}
\pspicture{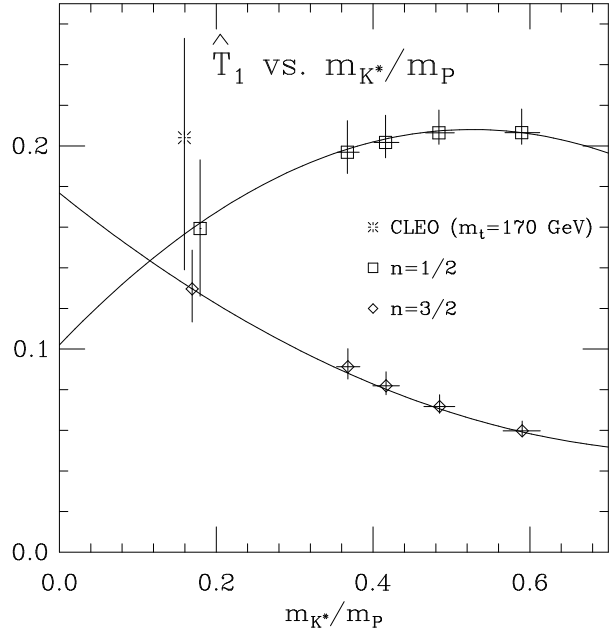}
\caption[]{${\hat T_1}$ extrapolation, for $N=1/2,3/2$ (Points at
$m_{K^*}/m_B$ displaced slightly for clarity). $\hat T_1$ is defined
in the text and agrees with $T_1(q^2{=}0)$ at $m_B$ for both
$N=1/2,3/2$.  The CLEO point is obtained from the CLEO measurement of
$BR(B \to K^*\gamma)$ as explained in the text.}
\label{figure:T1_hat_extrapolation}
\end{figure}

Linear and quadratic fits were carried out with the same functions as
for $\hat T_2$, allowing for correlations between masses and form
factors.  They are shown
in~Fig.(\ref{figure:T1_hat_extrapolation}). The $\chi^2/\mbox{d.o.f.}$
was approximately 1 for the $m_P^{3/2}$ scaling law, indicating that
the model is statistically valid in the available mass range. For the
$m_P^{1/2}$ scaling law we found a $\chi^2/\mbox{d.o.f.}$ of 0.3.

The correlated quadratic fit with radiative corrections gives,
\begin{equation}
\label{eq:T1_calc}
T_1(q^2{=}0;m_B;m_{K^*})  =
 \cases{{0.159}^{+34}_{-33} &$m_P^{1/2}$ scaling\cr
                    {0.124}^{+20}_{-18}&$m_P^{3/2}$ scaling\cr},
\end{equation}
where the errors quoted are statistical.

All methods of evaluating $T_1(q^2{=}0;m_P;m_{K^*})$ at
intermediate masses are compared in
table~\ref{table:T1_T2_comparison}. We consider the differences
between the methods as a measure of part of the systematic error. The
differences between the methods of determining the form factors at the
computed masses are of a similar size ($\sim 10\%$) to the systematic
error at the physical $B$ mass, as measured by the linear or quadratic
extrapolation of ${\hat T}_1$ in the inverse heavy meson mass.

The final result for $T_1(q^2{=}0;m_B;m_{K^*})$ is taken from the
quadratic fit for $T_1$, with an estimated systematic error in
extrapolation given by the difference between linear and quadratic
fits,
\begin{equation}
T_1(q^2{=}0;m_B;m_{K^*}) =
 \cases{{0.159}^{+34}_{-33}\pm{0.067} &$m_P^{1/2}$ scaling\cr
        {0.124}^{+20}_{-18}\pm{0.022}&$m_P^{3/2}$ scaling\cr}.
\end{equation}
The extrapolation is shown in
Fig.(\ref{figure:T1_hat_extrapolation}). We note that the value
obtained from $m_P^{3/2}$ scaling is consistent with the corresponding
value from $T_2$ calculated using the single pole $q^2$ behaviour
discussed earlier.

\subsection{$B_s \to \phi \gamma$}
Much of the analysis above can also be applied to the
rare decay $B_s \to \phi \gamma$.  ALEPH~\cite{aleph:btophigamma} and
DELPHI~\cite{delphi:btophigamma} have looked for this decay and
obtained 90\% CL upper bounds on its branching ratio of $4.1 \times
10^{-4}$ and $1.9 \times 10^{-3}$ respectively.  Future research into
this decay at LEP is planned.  The branching ratio for this decay can
be expressed in a form similar to~Eq.(\ref{eq:decay_rate}),
\begin{equation}
\label{eq:Bs-decay_rate}
\mbox{\it{BR\,}}(B_s \to \phi \gamma )
 = \frac{\alpha}{8 \pi^4} m_b^2 G_F^2
            m_{B_s}^3 \tau_{B_s} \left(1-\frac{m_{\phi}^2}{m_{B_s}^2}\right)^3
| V^{\phantom{*}}_{tb} V_{ts}^* |^2 |C_7(m_b)|^2 |T^s_1(q^2{=}0)|^2,
\end{equation}
where $T^s_1$ is the relevant form factor from the decomposition
of $\langle \phi | J_\mu | B_s \rangle$.
In determining this matrix element numerically, the interpolating
operator $J^V_\rho(x)$ is replaced by the operator $J^\phi_\rho(x)$
defined as,
\begin{equation}
J^\phi_\rho(x) = \overline{s}(x) \gamma_\rho s(x) .
\end{equation}
As a result of the presence of two identical particles in the final state,
there is an extra additive term in  the trace of~Eq.(\ref{eq:func-int-trace}),
which corresponds to $\overline{s} s$ creation from purely gluonic
states.  It is expected that this process is heavily suppressed by
Zweig's rule~\cite{zweig,okubo,iizuka}, and hence the extra term is
neglected.

As the variation of the form factors with respect to the
spectator quark mass has been discarded, it can be assumed that,
\begin{eqnarray}
T^s_1(q^2{=}0; m_P; m_\phi) &=& T_1(q^2{=}0; m_P; m_{K*}) , \\
T^s_2(q^2{=}0; m_P; m_\phi) &=& T_2(q^2{=}0; m_P; m_{K*}) .
\end{eqnarray}
By employing the same {\it ans\"atze\/} for extrapolating $T_1$ and
$T_2$ as the previous sections.
%
\begin{eqnarray}
T^s_1( q^2{=}0; m_{B_s}; m_\phi ) & = &
    \cases{0.165^{+32}_{-30}\pm{0.060}&$m_P^{1/2}$ scaling\cr
           0.125^{+20}_{-18}\pm{0.021}&$m_P^{3/2}$ scaling\cr}, \\
T^s_2( q^2_{max}; m_{B_s}; m_\phi ) & = &
             0.270^{+17}_{-9}\pm{0.009} , \\
T^{s, pole}_2( q^2{=}0; m_{B_s}; m_\phi ) & = &
           0.114^{+7}_{-4}{}^{+16}_{-15}.
\end{eqnarray}
We note that $T^s_1(q^2{=}0)$, with $m_P^{3/2}$ scaling, and
$T^{s,pole}_2(q^2{=}0)$ are consistent with each other.

\section{Conclusions}
In this paper we have reported on an {\it ab initio\/} computation of
the form factor for the decay $B \to K^* \gamma$.  The large number of
gauge configurations used in this calculation enables an extrapolation
to the appropriate masses to be made and gives a statistically
meaningful result.  To compare this result with experiment we convert
the preliminary branching ratio from CLEO, $\mbox{\it{BR\,}}(B \to
K^*\gamma) = (4.5 \pm 1.5 \pm 0.9) \times 10^{-5}$ based on 13 events
\cite{cleo:evidence-for-penguins}, into its corresponding $T_1$ form
factor, assuming the Standard Model.
We work at the scale $\mu=m_b=4.39\,\mbox{GeV}$, in the
$\overline{MS}$ scheme, using a pole mass of $M_b
=4.95(15)\,\mbox{GeV}$~\cite{morningstar:mb} to determine
$m_b$~\cite{broadhurst:pole-to-ms}.  Taking $|V_{ts}V_{tb}| =
0.037(3)$~\cite{stone:CKM}, $\tau_B = 1.5(2)
\,\mbox{ps}$~\cite{ALEPH:tau-b,OPAL:tau-B} and all other values from
the Particle Data Book combined with~Eq.(\ref{eq:decay_rate}), we find
$T^{\mbox{\scriptsize\it{exp}}}_1$ to be $0.23(6)$, $0.21(5)$ and
$0.19(5)$ for top quark masses of $m_t=100$, $150$ and
$200\,\mbox{GeV}$ respectively.  We find the calculated value for
$T_1$ consistent with these results to within two standard deviations.

In calculating the branching ratio, we use the perturbative
renormalisation of
$\sigma_{\mu\nu}$~\cite{borrelli:improved-operators} with a boosted
coupling, $g^2=1.7 g_0^2$, and the anomalous dimension, $\gamma_{{\bar
q}\sigma{q}} = -(8/3)(g^2/16\pi^2)$, to match the lattice results to
the continuum at the scale $\mu=m_b$, giving a matching coefficient of
$Z\approx0.95$. We apply a correction of $Z^2=0.90$ in the
calculations below.
Varying the scale of $C_7(\mu)$ from $\mu=m_b/2$ to $\mu=2 m_b$
changes the final branching ratio by $+27\%$ and $-20\%$ respectively.
This is due to the perturbative calculation of $C_7(\mu)$ and future
work on next-to-leading logarithmic order corrections will reduce
this variation significantly~\cite{buras:review}.

These uncertainties cancel in the dimensionless hadronisation ratio,
$R$,
\begin{eqnarray}
R &=& \frac{\mbox{\it{BR\,}}(B \to K^*\gamma)}
           {\mbox{\it{BR\,}}(B\to X_s \gamma)} \\
  &=& 4 {\left(\frac{m_B}{m_b}\right)}^3
      {\left(1-\frac{m^2_{K^\ast}}{m^2_B}\right) }^3
      |T_1(q^2{=}0)|^2,
\end{eqnarray}
which we find to be,
\begin{equation}
R=\cases{\left(
14.5^{+62}_{-60}\mbox{\,(stat.)\,}
\pm{6.1}\mbox{\,(sys.)\,}
\pm{1.6}\mbox{\,(exp.)\,}
\right)\% & $m_P^{1/2}$ scaling\cr
                        \left(
8.8^{+28}_{-25}\mbox{\,(stat.)\,}
\pm{3.0}\mbox{\,(sys.)\,}
\pm{1.0}\mbox{\,(exp.)\,}
\right)\% &$m_P^{3/2}$ scaling\cr}.
\end{equation}

Assuming the recent tentative result for $m_t$ from
CDF~\cite{fermilab_top_mass}, the lattice results give a branching
ratio for the decay $B \to K^*\gamma$ of,
\begin{equation}
\mbox{\it{BR\,}}(B \to K^*\gamma) = \cases{
\left(
2.5^{+11}_{-11}
\mbox{\,(stat.)\,}
\pm 2.1\mbox{\,(sys.)\,}
\pm 0.6\mbox{\,(exp.)\,}
{}^{+7}_{-5}\mbox{\,(scale)}
\right)
\times 10^{-5} &$m_P^{1/2}$ scaling\cr
                             \left(
1.5^{+5}_{-4}
\mbox{\,(stat.)\,}
\pm 0.5\mbox{\,(sys.)\,}
\pm 0.3\mbox{\,(exp.)\,}
{}^{+4}_{-3}\mbox{\,(scale)}
\right)
\times 10^{-5}
 &$m_P^{3/2}$ scaling\cr},
\end{equation}
where we separate the statistical and systematic errors from the
lattice, experimental and theoretical (scale) uncertainties. Combining
errors to produce an overall result yields,
\begin{equation}
\mbox{\it{BR\,}}(B \to K^*\gamma) = \cases{\left(
2.5
\pm1.3\mbox{\,(stat.)\,}
{}^{+28}_{-26} \mbox{\,(sys.)}
\right)
\times 10^{-5} &$m_P^{1/2}$ scaling\cr
                             \left(
1.5
\pm0.6\mbox{\,(stat.)\,}
{}^{+9}_{-8} \mbox{\,(sys.)}
\right)
\times 10^{-5}
&$m_P^{3/2}$ scaling\cr}.
\end{equation}
Similarly for $B_s \to \phi \gamma$, using
$m_{B_s}=5.3833(5)\,\mbox{GeV}$~\cite{OPAL:Bsmass,CDF:Bsmass} and
$\tau_{B_s}=1.54(15)\,\mbox{ps}$~\cite{forty:tau-Bs}, we find,
\begin{eqnarray}
\mbox{\it{BR\,}}(B_s \to \phi \gamma) &=&
\cases{\left(
2.8
{}^{+11}_{-10} \mbox{\,(stat.)\,}
\pm 2.1\mbox{\,(sys.)\,}
\pm 0.5\mbox{\,(exp.)\,}
{}^{+7}_{-5}\mbox{\,(scale)}
\right)
\times 10^{-5} &$m_P^{1/2}$ scaling\cr
       \left(
1.6
{}^{+5}_{-5} \mbox{\,(stat.)\,}
\pm 0.6\mbox{\,(sys.)\,}
\pm 0.3\mbox{\,(exp.)\,}
{}^{+4}_{-3}\mbox{\,(scale)}
\right)
\times 10^{-5}     &$m_P^{3/2}$ scaling\cr}, \\
                   &=&
\cases{\left(
2.8 \pm 1.2 \mbox{\,(stat.)\,}
{}^{+28}_{-26} \mbox{\,(sys.)}
\right)
 \times 10^{-5} &$m_P^{1/2}$ scaling\cr
       \left(
1.6 \pm 0.6 \mbox{\,(stat.)\,}
{}^{+10}_{-9} \mbox{\,(sys.)}
\right)
 \times 10^{-5}     &$m_P^{3/2}$ scaling\cr}.
\end{eqnarray}

In obtaining these results, we have made some assumptions.  Since this
calculation is carried out with one lattice spacing, we cannot explore
discretisation errors. However, the use of an $O(a)$-improved action
is expected to reduce these substantially.
As the form factors and mass ratios evaluated are dimensionless, we
also expect some of the systematic error from setting the scale to
cancel.
The extrapolation of matrix elements to the chiral limit has been
neglected, although the current data indicates a weak dependence on
the spectator quark mass.
Without doing a simulation using dynamical fermions, the error due to
quenching cannot be accurately estimated.  However, the good agreement
with experiment for other semileptonic, pseudoscalar to vector meson
decays~\cite{lubicz:d-decay-ii,bernard:d-decay}, that have been
determined using coarser lattices and lower statistics, suggests that
these errors are small.  We find our results consistent with previous
calculations~\cite{ukqcd:penguin-prl,bhs:penguin-prl}. With form
factors available over a range of masses, we have been able to
incorporate heavy-quark symmetry into our extrapolation and
investigate phenomenologically motivated pole-dominance models. These
methods supercede the simple linear extrapolation used as a guide in
our earlier preliminary study, where the limited set of two masses
precluded an investigation of different extrapolation
methods~\cite{ukqcd:penguin-prl}.

Whether pole dominance is a valid model for a large range of $q^2$ is
an important question. We have quoted results for two different
possibilities for the $q^2$ dependence of the form factors. Although
the lattice results visually favour $T_2$ constant in $q^2$, at least
for heavy quark masses around the charm mass, our fits favour a single
pole vector dominance form for $T_2$. The difference between the
results indicates the need for a better understanding of the combined
$q^2$ and heavy quark scaling behaviour of the relevant form factors.

We have not applied the constraint $T_1(0) = iT_2(0)$ to our fits in
this paper, using instead the consistency of our results with this
relation as a guide to the fitting method. We find that the the single
pole dominance model for the $q^2$ behaviour of $T_2$ (and
corresponding dipole behaviour for $T_1$) gives the most consistent
fit.  In this case we have attempted to determine the systematic
consistency by comparing $T_1(q^2{=}0;m_B;m_{K^*})$, extracted
using the $m_P^{3/2}$
scaling law, with $T_2(q^{2}{=}0;m_B;m_{K^*})$ assuming pole model
behaviour for $T_2$ and the expected pole mass.  It could be argued
that both methods are equivalent.  However, in extrapolating the form
factor $T_1(q^2{=}0;m_P;m_{K^*})$ to $m_B$, the coefficients in the
fit are not fixed, which is equivalent to letting the pole mass vary.
We require only that the leading order behaviour of $T_1$ satisfy the
$m_P^{3/2}$ dependence.

We look forward to improved experimental results for the decay
$B \to K^*\gamma$ and observation of $B_s \to \phi \gamma$. We hope future
lattice studies will significantly increase the accuracy of these
calculations.

\acknowledgments
The authors wish to thank G.~Martinelli for emphasising
the consistency requirements on scaling the form factors
$T_1$ and $T_2$.
They also thank
A.~Soni, C.~Bernard, A.~El-Khadra, and members of the UKQCD
collaboration,
including
C.~Allton, L.~Lellouch, J.~Nieves and H.~Wittig for useful discussions
on this topic.

JMF thanks the Nuffield Foundation for support under the scheme
of Awards for Newly Appointed Science Lecturers.
The University of Edinburgh and the Wingate Foundation
is acknowledged for its support of
HPS by a scholarship.  DGR (Advanced Fellow) and DSH (Personal Fellow)
acknowledge the support of the Science and Engineering Research
Council.
The authors acknowledge the support of the Particle Physics and Astronomy
Research Council by grant GR/J98202.

\begin{table}[htbp]
\begin{center}
\begin{tabular}{|c|c|c|c|c|c|c|}\hline
$\kappa_h$ & $\kappa_s$ & $\kappa_l$ & low $(qa)^2$ & high $(qa)^2$ &
 $ T_1(0)$ & $\chi^2/\mbox{d.o.f.}$ \\ \hline
0.12100 & 0.14144 & 0.14144 &$ -0.032^{+ 7}_{- 6} $&$ 0.258^{+ 7}_{- 9} $&
$ 0.283^{+ 19}_{- 12} $& 11.2/3 \\ \hline
0.12100 & 0.14144 & 0.14226 &$ -0.035^{+ 9}_{- 9} $&$ 0.265^{+ 11}_{- 12} $&
$ 0.282^{+ 30}_{- 21} $& 8.8/3 \\ \hline
0.12100 & 0.14144 & 0.14262 &$ -0.035^{+ 12}_{- 12} $&$ 0.271^{+ 15}_{- 16}
 $&$ 0.294^{+ 44}_{- 36} $& 4.1/3 \\ \hline
0.12100 & 0.14226 & 0.14144 &$ -0.014^{+ 10}_{- 9} $&$ 0.290^{+ 10}_{- 13} $&
$ 0.271^{+ 24}_{- 17} $& 7.3/3 \\ \hline
0.12100 & 0.14226 & 0.14226 &$ -0.022^{+ 16}_{- 15} $&$ 0.298^{+ 17}_{- 18} $&
$ 0.279^{+ 40}_{- 40} $& 6.7/3 \\ \hline
0.12100 & 0.14226 & 0.14262 &$ -0.028^{+ 22}_{- 22} $&$ 0.306^{+ 24}_{- 26} $&
$ 0.294^{+ 58}_{- 71} $& 4.2/3 \\ \hline
0.12500 & 0.14144 & 0.14144 &$ -0.118^{+ 5}_{- 5} $&$ 0.157^{+ 6}_{- 7} $&
$ 0.307^{+ 17}_{- 9} $& 10.3/3 \\ \hline
0.12500 & 0.14226 & 0.14144 &$ -0.104^{+ 8}_{- 7} $&$ 0.183^{+ 9}_{- 10} $&
$ 0.292^{+ 21}_{- 13} $& 8.5/3 \\ \hline
0.12900 & 0.14144 & 0.14144 &$ -0.188^{+ 4}_{- 4} $&$ 0.086^{+ 4}_{- 4} $&
$ 0.335^{+ 18}_{- 8} $& 9.4/3 \\ \hline
0.12900 & 0.14144 & 0.14226 &$ -0.190^{+ 6}_{- 5} $&$ 0.084^{+ 6}_{- 5} $&
$ 0.333^{+ 23}_{- 12} $& 7.2/3 \\ \hline
0.12900 & 0.14144 & 0.14262 &$ -0.190^{+ 8}_{- 6} $&$ 0.084^{+ 8}_{- 6} $&
$ 0.336^{+ 30}_{- 21} $& 3.2/3 \\ \hline
0.12900 & 0.14226 & 0.14144 &$ -0.177^{+ 6}_{- 5} $&$ 0.097^{+ 6}_{- 5} $&
$ 0.318^{+ 18}_{- 10} $& 7.1/3 \\ \hline
0.12900 & 0.14226 & 0.14226 &$ -0.182^{+ 10}_{- 8} $&$ 0.096^{+ 12}_{- 12} $&
$ 0.319^{+ 27}_{- 19} $& 7.5/3 \\ \hline
0.12900 & 0.14226 & 0.14262 &$ -0.186^{+ 14}_{- 12} $&$ 0.103^{+ 17}_{- 17} $&
$ 0.324^{+ 37}_{- 30} $& 3.6/3 \\ \hline
0.13300 & 0.14144 & 0.14144 &$ -0.240^{+ 3}_{- 2} $&$ 0.034^{+ 3}_{- 2} $&
$ 0.366^{+ 19}_{- 9} $& 7.3/3 \\ \hline
0.13300 & 0.14226 & 0.14144 &$ -0.233^{+ 4}_{- 3} $&$ 0.041^{+ 4}_{- 3} $&
$ 0.343^{+ 21}_{- 10} $& 5.7/3 \\ \hline
\end{tabular}
\end{center}

\caption{Results of pole fits to $T_1(q^2;m_P;m_V)$.}
\label{table:qsq_fits}
\end{table}
\begin{table}[htbp]
\begin{center}
\begin{tabular}{|c|c|c|c|c|c|c|}\hline
$\kappa_h$ & $\kappa_s$ & $\kappa_l$ & low $(qa)^2$ & high $(qa)^2$ &
 $ T_1(0)$ & $\chi^2/\mbox{d.o.f.}$ \\ \hline
0.12100 & 0.14144 & 0.14144 &$ -0.032^{+ 7}_{- 6} $&$ 0.258^{+ 7}_{- 9} $&
$ 0.279^{+ 20}_{- 12} $& 9.2/3 \\ \hline
0.12100 & 0.14144 & 0.14226 &$ -0.035^{+ 9}_{- 9} $&$ 0.265^{+ 11}_{- 12} $&
$ 0.275^{+ 32}_{- 24} $& 7.8/3 \\ \hline
0.12100 & 0.14144 & 0.14262 &$ -0.035^{+ 12}_{- 12} $&$ 0.271^{+ 15}_{- 16} $&
$ 0.286^{+ 46}_{- 36} $& 3.5/3 \\ \hline
0.12100 & 0.14226 & 0.14144 &$ -0.014^{+ 10}_{- 9} $&$ 0.290^{+ 10}_{- 13} $&
$ 0.266^{+ 26}_{- 19} $& 6.4/3 \\ \hline
0.12100 & 0.14226 & 0.14226 &$ -0.022^{+ 16}_{- 15} $&$ 0.298^{+ 17}_{- 18} $&
$ 0.274^{+ 43}_{- 42} $& 6.5/3 \\ \hline
0.12100 & 0.14226 & 0.14262 &$ -0.028^{+ 22}_{- 22} $&$ 0.306^{+ 24}_{- 26} $&
$ 0.286^{+ 67}_{- 66} $& 3.9/3 \\ \hline
0.12500 & 0.14144 & 0.14144 &$ -0.118^{+ 5}_{- 5} $&$ 0.157^{+ 6}_{- 7} $&
$ 0.308^{+ 18}_{- 9} $& 8.6/3 \\ \hline
0.12500 & 0.14226 & 0.14144 &$ -0.104^{+ 8}_{- 7} $&$ 0.183^{+ 9}_{- 10} $&
$ 0.291^{+ 21}_{- 14} $& 7.3/3 \\ \hline
0.12900 & 0.14144 & 0.14144 &$ -0.188^{+ 4}_{- 4} $&$ 0.086^{+ 4}_{- 4} $&
$ 0.337^{+ 18}_{- 8} $& 7.9/3 \\ \hline
0.12900 & 0.14144 & 0.14226 &$ -0.190^{+ 6}_{- 5} $&$ 0.084^{+ 6}_{- 5} $&
$ 0.334^{+ 23}_{- 12} $& 6.5/3 \\ \hline
0.12900 & 0.14144 & 0.14262 &$ -0.190^{+ 8}_{- 6} $&$ 0.084^{+ 8}_{- 6} $&
$ 0.337^{+ 30}_{- 20} $& 2.7/3 \\ \hline
0.12900 & 0.14226 & 0.14144 &$ -0.177^{+ 6}_{- 5} $&$ 0.097^{+ 6}_{- 5} $&
$ 0.319^{+ 18}_{- 10} $& 5.8/3 \\ \hline
0.12900 & 0.14226 & 0.14226 &$ -0.182^{+ 10}_{- 8} $&$ 0.096^{+ 12}_{- 12} $&
$ 0.320^{+ 27}_{- 19} $& 6.8/3 \\ \hline
0.12900 & 0.14226 & 0.14262 &$ -0.186^{+ 14}_{- 12} $&$ 0.103^{+ 17}_{- 17} $&
$ 0.325^{+ 37}_{- 30} $& 3.2/3 \\ \hline
0.13300 & 0.14144 & 0.14144 &$ -0.240^{+ 3}_{- 2} $&$ 0.034^{+ 3}_{- 2} $&
$ 0.362^{+ 19}_{- 9} $& 6.2/3 \\ \hline
0.13300 & 0.14226 & 0.14144 &$ -0.233^{+ 4}_{- 3} $&$ 0.041^{+ 4}_{- 3} $&
$ 0.341^{+ 20}_{- 10} $& 4.4/3 \\ \hline
\end{tabular}
\end{center}

\caption{Results of dipole fits to $T_1(q^2;m_P;m_V)$.}
\label{table:qsq_fits_b}
\end{table}
\begin{table}[htbp]
\begin{center}
\begin{tabular}{|c|c|c|c|c|c|c|}\hline
$\kappa_h$ & $\kappa_s$ & $\kappa_l$ & low $(qa)^2$ & high $(qa)^2$ &
 $ T_2(0)$ & $\chi^2/\mbox{d.o.f.}$ \\ \hline
0.12100 & 0.14144 & 0.14144 &$ -0.032^{+ 7}_{- 6} $&$ 0.289^{+ 6}_{- 7} $&
$ 0.301^{+ 22}_{- 12} $& 6.4/4 \\ \hline
0.12100 & 0.14144 & 0.14226 &$ -0.035^{+ 9}_{- 9} $&$ 0.293^{+ 9}_{- 10} $&
$ 0.310^{+ 32}_{- 23} $& 7.5/4 \\ \hline
0.12100 & 0.14144 & 0.14262 &$ -0.035^{+ 12}_{- 12} $&$ 0.297^{+ 13}_{- 15} $&
$ 0.315^{+ 44}_{- 39} $& 8.8/4 \\ \hline
0.12100 & 0.14226 & 0.14144 &$ -0.014^{+ 10}_{- 9} $&$ 0.318^{+ 9}_{- 11} $&
$ 0.288^{+ 25}_{- 17} $& 4.4/4 \\ \hline
0.12100 & 0.14226 & 0.14226 &$ -0.022^{+ 16}_{- 15} $&$ 0.324^{+ 15}_{- 16} $&
$ 0.300^{+ 38}_{- 31} $& 5.6/4 \\ \hline
0.12100 & 0.14226 & 0.14262 &$ -0.028^{+ 22}_{- 22} $&$ 0.330^{+ 22}_{- 25} $&
$ 0.312^{+ 52}_{- 67} $& 5.6/4 \\ \hline
0.12500 & 0.14144 & 0.14144 &$ -0.118^{+ 5}_{- 5} $&$ 0.190^{+ 5}_{- 6} $&
$ 0.322^{+ 19}_{- 10} $& 4.3/4 \\ \hline
0.12500 & 0.14226 & 0.14144 &$ -0.104^{+ 8}_{- 7} $&$ 0.214^{+ 8}_{- 9} $&
$ 0.310^{+ 22}_{- 13} $& 3.7/4 \\ \hline
0.12900 & 0.14144 & 0.14144 &$ -0.188^{+ 4}_{- 4} $&$ 0.108^{+ 3}_{- 4} $&
$ 0.351^{+ 17}_{- 8} $& 7.9/4 \\ \hline
0.12900 & 0.14144 & 0.14226 &$ -0.190^{+ 6}_{- 5} $&$ 0.109^{+ 6}_{- 6} $&
$ 0.353^{+ 23}_{- 12} $& 6.9/4 \\ \hline
0.12900 & 0.14144 & 0.14262 &$ -0.190^{+ 8}_{- 6} $&$ 0.112^{+ 8}_{- 9} $&
$ 0.349^{+ 30}_{- 20} $& 10.7/4 \\ \hline
0.12900 & 0.14226 & 0.14144 &$ -0.177^{+ 6}_{- 5} $&$ 0.126^{+ 6}_{- 7} $&
$ 0.332^{+ 19}_{- 10} $& 5.2/4 \\ \hline
0.12900 & 0.14226 & 0.14226 &$ -0.182^{+ 10}_{- 8} $&$ 0.129^{+ 9}_{- 10} $&
$ 0.337^{+ 28}_{- 20} $& 6.8/4 \\ \hline
0.12900 & 0.14226 & 0.14262 &$ -0.186^{+ 14}_{- 12} $&$ 0.133^{+ 14}_{- 15} $&
$ 0.336^{+ 36}_{- 34} $& 12.4/4 \\ \hline
0.13300 & 0.14144 & 0.14144 &$ -0.240^{+ 3}_{- 2} $&$ 0.045^{+ 2}_{- 3} $&
$ 0.372^{+ 14}_{- 6} $& 7.7/4 \\ \hline
0.13300 & 0.14226 & 0.14144 &$ -0.233^{+ 4}_{- 3} $&$ 0.057^{+ 4}_{- 5} $&
$ 0.351^{+ 16}_{- 7} $& 7.5/4 \\ \hline
\end{tabular}
\end{center}

\caption{Results of pole fits to $T_2(q^2;m_P;m_V)$.}
\label{table:T2_qsq_fits}
\end{table}
\begin{table}[htbp]
\begin{center}
\begin{tabular}{|c|c|c|c|c|c|c|}\hline
$\kappa_h$ & $\kappa_s$ & $\kappa_l$ & low $(qa)^2$ & high $(qa)^2$ &
 $ T_2(0)$ & $\chi^2/\mbox{d.o.f.}$ \\ \hline
0.12100 & 0.14144 & 0.14144 &$ -0.032^{+ 7}_{- 6} $&$ 0.289^{+ 6}_{- 7} $&
$ 0.344^{+ 15}_{- 5} $& 18.4/5 \\ \hline
0.12100 & 0.14144 & 0.14226 &$ -0.035^{+ 9}_{- 9} $&$ 0.293^{+ 9}_{- 10} $&
$ 0.334^{+ 19}_{- 8} $& 8.6/5 \\ \hline
0.12100 & 0.14144 & 0.14262 &$ -0.035^{+ 12}_{- 12} $&$ 0.297^{+ 13}_{- 15} $&
$ 0.322^{+ 23}_{- 13} $& 8.8/5 \\ \hline
0.12100 & 0.14226 & 0.14144 &$ -0.014^{+ 10}_{- 9} $&$ 0.318^{+ 9}_{- 11} $&
$ 0.323^{+ 16}_{- 6} $& 8.6/5 \\ \hline
0.12100 & 0.14226 & 0.14226 &$ -0.022^{+ 16}_{- 15} $&$ 0.324^{+ 15}_{- 16} $&
$ 0.323^{+ 22}_{- 10} $& 6.2/5 \\ \hline
0.12100 & 0.14226 & 0.14262 &$ -0.028^{+ 22}_{- 22} $&$ 0.330^{+ 22}_{- 25} $&
$ 0.314^{+ 28}_{- 17} $& 5.6/5 \\ \hline
0.12500 & 0.14144 & 0.14144 &$ -0.118^{+ 5}_{- 5} $&$ 0.190^{+ 5}_{- 6} $&
$ 0.353^{+ 15}_{- 6} $& 16.4/5 \\ \hline
0.12500 & 0.14226 & 0.14144 &$ -0.104^{+ 8}_{- 7} $&$ 0.214^{+ 8}_{- 9} $&
$ 0.333^{+ 16}_{- 6} $& 6.6/5 \\ \hline
0.12900 & 0.14144 & 0.14144 &$ -0.188^{+ 4}_{- 4} $&$ 0.108^{+ 3}_{- 4} $&
$ 0.365^{+ 15}_{- 5} $& 12.6/5 \\ \hline
0.12900 & 0.14144 & 0.14226 &$ -0.190^{+ 6}_{- 5} $&$ 0.109^{+ 6}_{- 6} $&
$ 0.361^{+ 20}_{- 8} $& 7.5/5 \\ \hline
0.12900 & 0.14144 & 0.14262 &$ -0.190^{+ 8}_{- 6} $&$ 0.112^{+ 8}_{- 9} $&
$ 0.353^{+ 24}_{- 12} $& 10.8/5 \\ \hline
0.12900 & 0.14226 & 0.14144 &$ -0.177^{+ 6}_{- 5} $&$ 0.126^{+ 6}_{- 7} $&
$ 0.342^{+ 15}_{- 5} $& 6.2/5 \\ \hline
0.12900 & 0.14226 & 0.14226 &$ -0.182^{+ 10}_{- 8} $&$ 0.129^{+ 9}_{- 10} $&
$ 0.339^{+ 21}_{- 9} $& 6.9/5 \\ \hline
0.12900 & 0.14226 & 0.14262 &$ -0.186^{+ 14}_{- 12} $&$ 0.133^{+ 14}_{- 15} $&
$ 0.335^{+ 30}_{- 16} $& 12.4/5 \\ \hline
0.13300 & 0.14144 & 0.14144 &$ -0.240^{+ 3}_{- 2} $&$ 0.045^{+ 2}_{- 3} $&
$ 0.374^{+ 14}_{- 6} $& 8.5/5 \\ \hline
0.13300 & 0.14226 & 0.14144 &$ -0.233^{+ 4}_{- 3} $&$ 0.057^{+ 4}_{- 5} $&
$ 0.351^{+ 15}_{- 5} $& 7.5/5 \\ \hline
\end{tabular}
\end{center}

\caption{Results of constant fits to $T_2(q^2;m_P;m_V)$.}
\label{table:T2_qsq_fits_b}
\end{table}
\begin{table}[htbp]
\begin{center}
\begin{tabular}{|c|c|c|c|c|c|c|}\hline
\multicolumn{2}{|c|}{ } & \multicolumn{2}{c|}{$T_1(m_q) = a + b m_q$}
& 	& $T_1(m_q) = c$ &  \\ \hline
$\kappa_h$ & $\kappa_s$ & $a$ & $b$ & $\chi^2/\mbox{d.o.f.}$ & $c$ &
$\chi^2/\mbox{d.o.f.}$ \\ \hline
0.121 &	0.14144 &$ 0.299^{+ 40}_{- 33} $&$ -0.362^{+  677}_{-  626} $& 0.04/1
&$ 0.281^{+ 18}_{- 12}   $& 0.3/2 \\ \hline
0.121 & 0.14226 &$ 0.289^{+ 59}_{- 54} $&$ -0.326^{+  100}_{-  104} $& 0.05/1
&$ 0.271^{+ 21}_{- 12} $& 0.1/2 \\ \hline
0.121 & 0.1419 &$ 0.293^{+ 49}_{- 42} $&$ $&  &$ 0.275^{+ 19}_{- 12} $
&  \\ \hline
0.129 & 0.14144 &$ 0.336^{+ 27}_{- 16} $&$ -0.058^{+  352}_{-  391} $&
0.1/1 &$ 0.333^{+ 18}_{- 9}  $& 0.1/2 \\ \hline
0.129 & 0.14226 &$ 0.330^{+ 30}_{- 21} $&$ -0.288^{+  403}_{-  413} $&
0.2/1 &$ 0.316^{+ 17}_{- 9} $& 0.8/2 \\ \hline
0.129 & 0.1419 &$ 0.333^{+ 27}_{- 18} $&$ $& &$ 0.324^{+ 17}_{- 9} $
&  \\ \hline
\end{tabular}
\end{center}
\caption{Extrapolation of $T_1(q^2=0)$, from a single pole fit, to the
chiral limit, where $T_1$ is assumed either to have a linear
dependence on the pole mass of the light quark, or to be independent
of the pole mass.
$\kappa_{\protect\mbox{\protect\scriptsize\protect\it{strange}}}
= 0.1419$ corresponds to the physical strange quark mass from
determining the mass of the $K$ on this lattice.}
\label{table:Tone_chiral}
\end{table}
\begin{table}[htbp]
\begin{center}
\begin{tabular}{|c|c|c|c|c|c|c|}\hline
\multicolumn{2}{|c|}{ } & \multicolumn{2}{c|}{$T_2(q^2{=}0;m_q) = a + b m_q$}
& 	& $T_2(q^2{=}0;m_q) = c$ &  \\ \hline
$\kappa_h$ & $\kappa_s$ & $a$ & $b$ & $\chi^2/\mbox{d.o.f.}$ & $c$ &
$\chi^2/\mbox{d.o.f.}$ \\ \hline
0.12900 & 0.14144 &$  0.363^{+  22}_{-  24} $&$
-0.311^{+  460}_{-  394} $&   0.1/1 &$
0.348^{+  13}_{-  9} $&   0.6/2 \\ \hline
0.12900 & 0.14226 &$  0.337^{+  23}_{-  31} $&$
-0.190^{+  713}_{-  396} $&   0.1/1 &$
0.329^{+  17}_{-  9} $& 0.3/2 \\ \hline
0.12900 &  0.14190 &$  0.349^{+  22}_{-  27} $&$
 $& &$ 0.337^{+  15}_{-  9} $&  \\ \hline

0.12100 & 0.14144 &$  0.323^{+  32}_{-  43} $&$
-0.406^{+  826}_{-  596} $&   0.01/1 &$
0.304^{+  17}_{-  14} $& 0.3/2 \\ \hline
0.12100 & 0.14226 &$  0.311^{+  23}_{-  47} $&$
-0.531^{+  1008}_{-  452} $&   0.1/1 &$
0.289^{+  19}_{-  15} $& 0.6/2 \\ \hline
0.12100 &  0.14190 &$  0.317^{+  25}_{-  42} $&$
$& &$ 0.296^{+  18}_{-  15} $& \\ \hline
\end{tabular}
\end{center}
\caption{Extrapolation of $T_2(q^2{=}0)$, from a single pole fit,
to the chiral limit, where $T_2$ is assumed either to have a linear
dependence on the pole mass of the light quark, or to be independent
of the pole
mass. $\kappa_{\protect\mbox{\protect\scriptsize\protect\it{strange}}}
= 0.1419$ corresponds to the physical strange quark mass from
determining the mass of the $K$ on this lattice.}
\label{table:Two_inter_chiral}
\end{table}
\begin{table}[htbp]
\begin{center}
\begin{tabular}{|c|c|c|c|c|c|c|}\hline
\multicolumn{2}{|c|}{ } & \multicolumn{2}{c|}{$T_2(q^2_{max};m_q) = a + b m_q$}
& 	& $T_2(q^2_{max};m_q) = c$ &  \\ \hline
$\kappa_h$ & $\kappa_s$ & $a$ & $b$ & $\chi^2/\mbox{d.o.f.}$ & $c$ &
$\chi^2/\mbox{d.o.f.}$ \\ \hline
0.12100 & 0.14144 &$  0.331^{+  29}_{-  14} $&$
0.492^{+  348}_{-  434} $&   0.4/1 &$
0.353^{+  15}_{-  7} $& 1.9/2 \\ \hline
0.12100 & 0.14226 &$  0.327^{+  31}_{-  17} $&$
-0.026^{+  378}_{-  466} $&   1.9/1 &$
0.325^{+  15}_{-  6} $&   1.9/2 \\ \hline
0.12100 &  0.14190 &$ 0.328^{+  29}_{-  16} $&$ $& &$
 0.337^{+ 15}_{-  6} $&
 \\ \hline
0.12900 & 0.14144 &$  0.370^{+  27}_{-  10} $&$
-0.136^{+  198}_{-  368} $&   0.9/1 &$
0.363^{+  13}_{-  6} $&   1.1/2 \\ \hline
0.12900 & 0.14226 &$  0.349^{+  32}_{-  16} $&$
-0.136^{+  308}_{-  453} $&   0.7/1 &$
0.341^{+  13}_{-  6} $& 0.8/2 \\ \hline
0.12900 &  0.14190 &$  0.358^{+  29}_{-  13} $&$
 $&  &$
 0.351^{+  13}_{- 6} $&
  \\ \hline
\end{tabular}
\end{center}
\caption{Extrapolation of $T_2(q^2_{max})$ to the chiral limit, where $T_2$
is assumed either to have a linear dependence on the pole mass of the
light quark, or to be independent of the pole
mass. $\kappa_{\protect\mbox{\protect\scriptsize\protect\it{strange}}}
= 0.1419$ corresponds to the physical strange quark mass from
determining the mass of the $K$ on this lattice.}
\label{table:Two_chiral}
\end{table}
\begin{table}[htbp]
\begin{center}
\begin{tabular}{|c|c|c|c|c|c|c|c|c|}\hline
$\kappa_h$ & $m_{K^*}/m_P$ & $T_2(q^2_{max})$ & $q^2_{max}/m^2_{P_{s1}}$ &
 $T_2(0)$ {\scriptsize(a)} &
 $T_1(0)$ {\scriptsize(b)} &
 $T_1(0)$ {\scriptsize(c)} &
 $T_2(0)$ {\scriptsize(d)} &
 $T_2(0)$ {\scriptsize(e)} \\ \hline
0.13300 &$ 0.59^{+ 2}_{- 2} $&$ 0.362^{+ 15}_{- 6} $&$ 0.08^{+ 1}_{- 1} $&
$ 0.333^{+ 14}_{- 7} $&$ 0.356^{+ 19}_{- 10} $&$ 0.353^{+ 19}_{- 8} $&
$ 0.359^{+ 15}_{- 7} $&$ 0.361^{+ 14}_{- 4}$\\ \hline
0.12900 &$ 0.48^{+ 2}_{- 2} $&$ 0.353^{+ 15}_{- 6} $&$ 0.15^{+ 1}_{- 1} $&
$ 0.301^{+ 14}_{- 7} $&$ 0.324^{+ 19}_{- 9}  $&$ 0.326^{+ 17}_{- 7} $&
$ 0.339^{+ 17}_{- 9} $&$ 0.352^{+ 15}_{- 4} $\\ \hline
0.12500 &$ 0.42^{+ 2}_{- 2} $&$ 0.346^{+ 16}_{- 6} $&$ 0.20^{+ 1}_{- 1} $&
$ 0.276^{+ 14}_{- 7} $&$ 0.298^{+ 19}_{- 11} $&$ 0.298^{+ 19}_{- 10} $&
$0.318^{+ 20}_{- 12} $&$ 0.342^{+ 15}_{- 5} $\\ \hline
0.12100 &$ 0.37^{+ 2}_{- 1} $&$ 0.339^{+ 16}_{- 7} $&$ 0.26^{+ 2}_{- 1} $&
$ 0.252^{+ 14}_{- 8} $&$ 0.278^{+ 22}_{- 14} $&$ 0.276^{+ 22}_{- 13} $&
$ 0.298^{+ 23}_{- 15}$&$ 0.332^{+ 15}_{- 4} $\\ \hline
	 &$ 0.1692^{+ 1}_{- 1} $&$ 0.269^{+ 17}_{- 9} $&$ 0.51^{+ 6}_{- 6} $&
$ 0.112^{+ 7}_{- 7} $&$ 0.124^{+ 20}_{- 18} $&$  0.159^{+ 34}_{- 33} $&
&  \\ \hline
\end{tabular}
\end{center}
\caption{Comparison of results from different methods of
extracting $T_{1,2}(q^2{=}0)$.  The last row indicates the final
extrapolation to the physical regime $m_{K^*}/m_B$, with results of
the three methods of extracting $T_{1,2}(q^2{=0};m_B;m_{K^*})$ at the
feet of columns labelled (a), (b) and (c). Column labels:
(a) pole form, with $1^+$ mass from two-point functions [estimated
{}from $B_{s1}$ mass for final extrapolation],
(b) dipole form, with $m_P^{3/2}$ scaling for final extrapolation,
(c) pole form, with $m_P^{1/2}$ scaling for final extrapolation,
(d) pole form, with mass determined from pole fit,
(e) constant form factor.}
\label{table:T1_T2_comparison}
\end{table}

%
%
%

\end{document}